\documentclass[12pt]{article}

\usepackage{bbold}
\usepackage{amssymb,amsmath,bbm,cite}
\usepackage{footnpag} 
\usepackage[all]{xy}
\usepackage{graphics}

\usepackage{graphicx}

\catcode`\@=11

\@addtoreset{equation}{section}
\def\theequation{\arabic{section}.\arabic{equation}}
     
\catcode`\@=11
\def\thesection{\arabic{section}}

\def\appendix{\setcounter{section}{0}
        \def\thesection{Appendix}
        \def\theequation{\Alph{section}.\arabic{equation}}}
\def\section{\@startsection{section}{1}{\z@}{3.5ex plus 1ex minus
   .2ex}{2.3ex plus .2ex}{\large\bf}}

\newcommand{\captionfonts}{\footnotesize}
\makeatletter  
\long\def\@makecaption#1#2{%
  \vskip\abovecaptionskip
  \sbox\@tempboxa{{\captionfonts #1: #2}}%
  \ifdim \wd\@tempboxa >\hsize
    {\captionfonts #1: #2\par}
  \else
    \hbox to\hsize{\hfil\box\@tempboxa\hfil}%
  \fi
  \vskip\belowcaptionskip}
\makeatother   

\long\def\@makefntext#1{\parindent 0cm\noindent
\hbox to 1em{\hss$^{\@thefnmark}$}#1}

\def\wt#1{\widetilde{#1}}

\def\d{\partial}

\def\be{\begin{equation}}
\def\ee{\end{equation}}
\def\beq{\begin{equation}}
\def\eeq{\end{equation}}
\def\bea{\begin{eqnarray}}
\def\eea{\end{eqnarray}} 
\def\beqa{\begin{equation}\begin{array}{l}}
\def\eeqa{\end{array}\end{equation}}

\def\eqn#1{(\ref{#1})}

\def\eqref#1{eq.~(\ref{eq:#1})}

\def\half{\mbox{\small{$\frac{1}{2}$}}}

\def\nn{\nonumber}

\newcommand{\Chr}[3]{\Gamma^{#1}_{#2#3}}

\newcommand{\tmg}{topologically massive gravity}

\newcommand{\SL}{{\rm SL}}

\newcommand{\ssp}{{\scriptscriptstyle +}}
\newcommand{\ssm}{{\scriptscriptstyle -}}
\newcommand{\sspm}{{\scriptscriptstyle \pm}}

\def\RR{\mathbb R}

\def\om{\omega}

\begin{document}


\vspace{.8cm}
\setcounter{footnote}{0}
\begin{center}
{\Large{\bf 
Cosmological Topologically Massive\\[2mm] Gravitons and Photons\footnote{Some of these results were reported in abbreviated form in \cite{letter}.}}
    }\\[10mm]

{\sc S.\ Carlip$^*$, S.\ Deser$^\ddag$,
A.\ Waldron$^\dag$, and D.\ K.\ Wise$^\dag$
\\[6mm]}

{\em\small  
$^*$Department of Physics,
University of California, Davis, CA 95616,
USA\\ {\tt carlip@physics.ucdavis.edu}}\\[5mm]
{\em\small  
$^\ddag$Lauritsen Laboratory, California Institute of Technology, Pasadena, CA 91125 and\\
Department of Physics, Brandeis University, Waltham,
MA 02454, 
USA\\ {\tt deser@brandeis.edu}}\\[5mm]
{\em\small  
$^\dag$Department of Mathematics,
University of California, Davis, CA 95616,
USA\\ {\tt wally,derek@math.ucdavis.edu}}\\[5mm]

{\small\sc June 18, 2008}

\bigskip

\bigskip

{\sc Abstract}\\
\end{center}

{\small
\begin{quote}

We study topologically massive (2+1)-dimensional gravity with a negative
cosmological constant.  The masses of the linearized curvature excitations about 
AdS$_3$ backgrounds are not only shifted from their flat background values but, more 
surprisingly, split according to chirality.  For all finite values of the topological mass, 
we find a single bulk degree of freedom with positive energy, and exhibit a complete
set of normalizable, finite-energy wave packet solutions.  This model can also be written 
as a sum of two higher-derivative $\SL(2,\RR)$ Chern--Simons theories, weighted by 
the central charges of the boundary conformal field theory.  At two particular `critical' 
values of the couplings, one of these central charges vanishes, and linearized 
topologically massive gravity becomes equivalent to topologically massive 
electromagnetism; however, the physics of the bulk wave packets remains unaltered 
here.




\end{quote}
}

\newpage


\section{Introduction}

Topologically massive theories~\cite{Deser:1982vy} are three-dimensional gauge 
invariant systems with actions consisting of the normal kinetic term plus a Chern-Simons 
term, and come in both vector (abelian or Yang--Mills) and tensor (gravitational) 
versions.  These theories' excitations were found to have rather surprising 
complementary properties: they are both massive and gauge invariant, and their
single, parity-violating, degrees of freedom are represented by indexless 
``scalar'' fields with nonvanishing spins.  Further, the energy of the linearized 
gravitons is manifestly positive if and only if the sign of the Einstein term is taken 
to be the opposite of the usual one in $d=4$. 

While these results are quite old, 
the obvious cosmological completion of topologically massive gravity---the
addition of a cosmological constant---was only studied more 
recently~\cite{Deser:2003vh,Kraus,Solodukhin,Strom}.  However, its supersymmetric 
(supergravity) extension was given long ago~\cite{Deser:1984py}, and has the 
powerful corollary  that the underlying bosonic model (with the above sign!) has 
positive energy---which is still defined in an anti-de Sitter context~\cite{Abbott:1981ff}---because 
of the usual $E=\{Q,Q^\dagger\}$ SUSY relation.   (As in all 
supergravities, a negative cosmological constant is required.)  Similar considerations hold 
for the graviton's spin~3/2 partner~\cite{3/2}.   Cosmological topologically massive 
gravity is the system we will study here, with a warm-up via scalars and cosmological 
topologically massive electrodynamics.

As is well known, fields in (anti-) de Sitter backgrounds can behave quite differently from 
those in flat spacetime.   Indeed, some of our results will be reminiscent of those obtained 
earlier~\cite{Deser:1983tm,Deser:2001pe} for ``partially massive'' tensor models in $d=4$. 
The  effective masses of our excitations are shifted, but a suitable tuning in the 
$(m^2, \Lambda)$ plane  simultaneously keeps a single degree of freedom while
allowing propagation on the null cone. Importantly, what remains unchanged is the single
degree of freedom and its description via an indexless ``scalar'' field. This field 
has a mass that always respects the Breitenlohner--Freedman bound, which allows
an extended range of negative squared-masses $m^2\geq \Lambda$ for fields in anti-de Sitter space
\cite{Breitenlohner:1982jf,Mezincescu:1984ev}.
It is thus a consistent field theory.  We demonstrate this 
result both in light-front gauge and in terms of manifestly gauge-invariant variables. 

Perhaps even more interesting than the possibility of lightlike propagation for special 
tunings of the mass parameter is the chirality dependence of the mode masses.  
The decay rate of metric fluctuations near the anti-de Sitter boundary also depends on their 
chirality.   This has consequences for the  discussion of boundary behavior in~\cite{Strom} 
and the conjecture that the bulk modes disappear and the boundary theory
becomes chiral  at a critical value~$\mu^2= -\Lambda$ of the `topological mass' $\mu$.   We find that at this
critical point, the metric fluctuations can form finite-energy wave packets, with finite
 Fefferman--Graham asymptotics \cite{Fefferman} and
no constraints on (three-dimensional) chirality.\footnote{There is an unfortunate semantic 
confusion coming from two different meanings of the term ``chiral.''   We are using the
term here in the three-dimensional sense: our bulk modes depend on both $x^+$ and
$x^-$ in light-front coordinates.  Similarly, the solutions we discuss in section
\ref{chirspec} are chiral in the sense that they depend, in the bulk, only on $x^+$ and 
not $x^-$.   This is quite distinct from the boundary chirality discussed in \cite{Strom,Stromc}, 
which is determined by global properties of the diffeomorphism generators.}  
For all smaller values of the mass parameter, 
however, the Fefferman--Graham expansion contains divergent terms, perhaps necessitating a 
truncation to a single handedness or addition of boundary counterterms to the action.
Our results show that the mechanism behind this truncation is not a new gauge invariance.  
We stress that for generic masses---including the critical value---the theory describes a 
single, positive energy bulk mode with correct asymptotics.  These results connect smoothly 
to existing Minkowski ones~\cite{Deser:1982vy} as the cosmological constant is taken to 
zero.  There also exist exact pp-wave chiral solutions at the critical point and at higher 
values of $|\mu|$.

As noted above, positive energy for the massive fluctuations requires the 
Einstein part of the action to have  
a sign opposite  to the canonical choice for pure ($d=4$) gravity.  This leads to no 
inconsistencies in flat vacua, since three-dimensional pure Einstein massless bulk gravitational modes 
are pure gauge.  For a negative cosmological constant, however, 
this choice of sign also leads to negative energy for the BTZ black hole, which is an 
exact solution not only of pure Einstein gravity, but also of our field equations for 
arbitrary~$\mu$~\cite{Moussa}.  This 
suggests that the theory may be fundamentally unstable.  While this is a serious concern,
instability is by no means certain:  Classically, it is not clear that a negative-energy
black hole can be created from positive-energy matter, and while the potential quantum 
instability may pose a more serious problem, we know of no instanton that mediates the 
production of negative-energy black holes.  Indeed, the 
possibility remains that one can find a superselection sector in which BTZ black holes
are excluded, much as one excludes negative-energy Schwarzschild black holes in 3+1 
dimensions. The supergravity positive energy arguments of~\cite{Deser:1984py} provide
corroborative evidence.

The alternative considered in~\cite{Strom} is to flip the overall sign of the action, 
keeping positive energy BTZ black holes, but at the price of  negative-energy bulk excitations.  
There it was argued  
that at the critical value $\mu^2= -\Lambda$, the propagating bulk degrees of freedom become 
pure gauge, restoring the consistency of the theory.  We fail to find this behavior; rather,
we exhibit well-behaved propagating bulk modes with good asymptotic properties
(but negative energy with the flipped sign of the action).

The critical value of $\mu$ does have its own peculiarities, however.   
Like three-dimensional Einstein gravity~\cite{AT,Witten3dgrav}, the theory can be 
written for any $\mu$ as a sum of two (higher-derivative) Chern--Simons theories.
At $\mu^2= -\Lambda$, this action degenerates to a single such Chern--Simons term.  Separately, 
we demonstrate that  linearized topologically massive gravity at this critical point is 
equivalent to topologically massive electrodynamics in an anti-de Sitter background.  
This again confirms that  linearized topologically massive gravity describes a single, 
consistent field theoretic bulk degree of freedom, also at this point.

We will begin by reviewing  scalar fields in AdS$_3$,
focusing on the light-front formalism.  We next discuss topologically massive photons in 
Section~\ref{gamma}, reducing the dynamics to a single ``scalar'' mode in a light-front 
approach.  We then rederive the same results in terms of manifestly gauge invariant 
curvatures, and display the splitting of masses according to the chirality of the excitation.  
In Section~\ref{gravitons}, we apply the same logic to linearized metric fluctuations, and 
show that cosmological massive gravitons can be described as ``scalars,'' again with masses 
that depend on their chirality.  All of our masses satisfy the three-dimensional 
Breitenlohner--Freedman bounds~\cite{Breitenlohner:1982jf,Mezincescu:1984ev}, 
and therefore describe positive energy excitations.   In the next two sections, we study the 
asymptotics of these metric fluctuations and the construction of a complete set of normalizable, 
finite-energy wave packets.  In Sections~\ref{CSG} and~\ref{gravgamma}, we explore 
the Chern--Simons action at the critical topological mass, and reformulate the linearized 
graviton theory as topologically massive electrodynamics.  In Section~\ref{chirspec}, 
we examine the chiral solutions and show that they are exact pp-waves.    
Finally, many detailed but useful calculations are relegated to an Appendix.  These include 
explicit solutions for all of the bulk excitations, the scalar bulk-boundary intertwiner, 
and a simple rederivation of the Breitenlohner--Freedman bound.

\section{AdS$_3$ Scalars}
\label{HS}

We denote by $\gamma_{\mu\nu}$ the dynamical metric, with signature $({-}{+}{+})$, 
reserving~$g_{\mu\nu}$ for the background AdS$_3$ metric and $D$ for the background 
connection.  Our other conventions are as follows: 
the Ricci tensor---equivalent in three dimensions to the full curvature---is
\[
      R_{\nu\tau}\equiv{R^\sigma}_{\nu\sigma\tau} 
      = \d_\sigma \Chr{\sigma}{\tau}{\nu}-\d_\tau \Chr{\sigma}{\sigma}{\nu} 
       + \Chr{\sigma}{\sigma}{\mu}\Chr{\mu }{\tau}{\nu} 
       - \Chr{\sigma}{\tau}{\mu}\Chr{\mu}{\sigma}{\nu}\, ,
\]
and the scalar curvature is the positive trace, $R = {R^\mu}_{\mu}$.  In particular, for 
the AdS$_3$ background ($\Lambda<0$), we have
\be
R_{\alpha \beta\mu\nu} 
  = \Lambda(g_{\alpha\mu}g_{\beta\nu} - g_{\alpha\nu}g_{\beta\mu})\, , \quad
R_{\mu\nu}=2\Lambda g_{\mu\nu} \quad {\rm and} \quad R = 6\Lambda.
\ee
We employ units  
\be
\Lambda=-1\, ;
\ee
the cosmological constant can always be reinstated by 
dimensional analysis.

In our primarily light-front approach, we use the Poincar\'e frame
\be
ds^2 = g_{\mu\nu} dx^\mu dx^\nu = \frac{2dx^+ dx^- + dz^2}{z^2}\, ,\label{Poincare}
\ee
where $x^\pm = \frac{1}{\sqrt{2}}(x\pm t)$.  We view $x^+\equiv \tau$ as a time coordinate 
and often denote $\frac\d{\d\tau}$ by a dot or $\partial_+$.  We also often write $\partial$ for 
$\frac \d{\d z}$ and $\partial_-$ for $\frac\d{\d x^-}$.   

While these choices greatly simplify many computations, the coordinates~(\ref{Poincare}) do not 
cover the whole of anti-de Sitter space.  Recall that AdS$_3$ is the 
simply connected covering space of the
hyperboloid $-U^2-V^2+X^2+Y^2=-1$.   A Poincar\'e patch covers half of this hyperboloid, 
say the region $U+X>0$, and the coordinates~(\ref{Poincare})
correspond to setting 
$U+X=1/z$, $Y=x/z$, $V=-t/z$, $U-X=(-t^2+x^2+z^2)/z$.  Results 
in other Poincar\'e patches are easily obtained by successive inversion operations:
for example, the patch $X-U>0$ is reached by the diffeomorphism 
$x^\mu=-\wt x^\mu/(\wt x^\nu\eta_{\nu\rho} \wt x^\rho)$, under which the form
of the 
metric is invariant.  

\begin{figure}
\[
\xy
(0,0)*{\includegraphics[height=5cm]{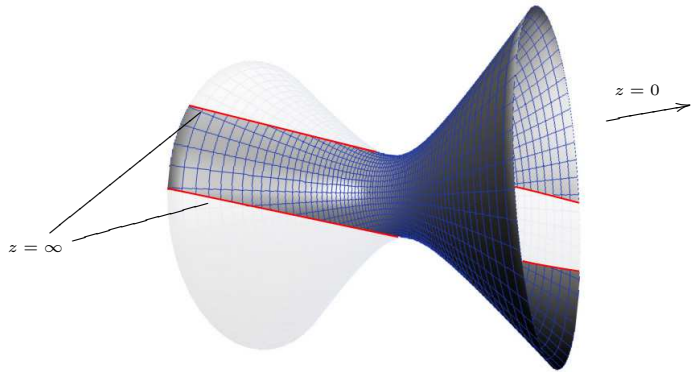}};
(-45,-8)="A"*{\tiny \txt{$z=\infty$}};
(35,13)="B"*{\txt{\tiny $z=0$}};
"A"+(2,2);(-23,10)**\crv{};
"A"+(5,1);(-22,-2)**\crv{};
"B"+(-4,-4);"B"+(7,-2)**\crv{}*\dir{>};
\endxy
\]
\caption{\footnotesize A 2d cross-section---corresponding to $V = 0$ in the ambient 
four-dimensional 
coordinates, or equivalently $t=0$ in Poincar\'e coordinates---of the Poincar\'e patch, 
showing how it sits in the hyperboloid model of AdS$_3$.  The surface $z=0$ lies in the 
AdS$_3$ boundary.  The $z=\infty$ surface intersects any constant-Poincar\'e-time slice 
in a pair of lines through the bulk, as shown; these meet asymptotically (to the left, in 
the picture) in a single point on the boundary. \label{patch}}
\end{figure}

The key idea of the light-front method is that actions are automatically first order, albeit 
with a nonstandard symplectic form given by the anti-symmetric bracket
\be
\langle A,B \rangle \equiv \int dx^-dz\;  A \partial_- B = -\langle B,A\rangle\, . 
\ee
In particular, the standard (positive energy) massive scalar field action
\be
I=-\frac 12 \int d^3x \sqrt{-g}\, \Big\{\partial_\mu \varphi g^{\mu\nu} 
  \partial_\nu\varphi + m^2 \varphi^2\Big\}\, ,
\ee
rewritten in light-front coordinates, takes the Hamiltonian form
\be
\label{scalaraction}
I=\int d\tau \left\{\langle \phi,\dot\phi\rangle - \left(\frac12 [\partial \phi]^2
+\frac1{2z^2}\left[m^2 + \frac 34\right]\phi^2\right)dx^-dz  \right\}\, ,
\ee
where we have made the field redefinition
\be
\phi \equiv \frac{1}{\sqrt{z}}\varphi\,  .
\label{scale0}
\ee
Varying this action yields the equation of motion
\be
\Big[2\partial_-\partial_+  + \partial^2 - \frac{m^2+3/4}{z^2}\Big]\phi = 0\, . 
\label{scalareom}
\ee
Setting $m^2 = -3/4$ yields the massless Minkowski
wave equation, which is hardly surprising, as the background is conformally
flat, and moreover this value corresponds precisely to adding the conformal improvement 
term~$\frac{1}{16} R \varphi^2$ to the scalar action, which gives lightlike 
propagation~\cite{Deser:1983tm}.  The scaling (\ref{scale0}) is of course the usual 
conformal one for a scalar in the metric (\ref{Poincare}).  Note also that this negative value of $m^2$ is 
consistent, since the Breitenlohner--Freedman bound in three dimensions allows 
for masses as low as $m^2 = -1$~\cite{Breitenlohner:1982jf}.

For generic values of the mass, this equation is most easily solved by going to Fourier 
space in the $x^\pm$ coordinates.  The expression  $2\partial_-\partial_+ = -\partial_t^2 
+ \partial_x^2$ then transforms to the frequency  $\om^2 = E^2 - p_x^2$, yielding Bessel's 
equation in the variable $z$ for the combination $\varphi(z)/z$,
\be
\label{bessel}
\Big[\frac{d^2}{dz^2}+\frac1z\frac{d}{dz}
  +\omega^2-\frac{\nu^2}{z^2}\Big]\Big(\frac{\varphi}{z}\Big)=0\, ,
\ee
 with index 
 \be
 \nu^2 = m^2 +1\, .
 \ee
The solutions of (\ref{bessel}) are oscillatory; indeed, when $\nu$ is a half-integer, the 
Bessel function solution reduces to a slowly varying function of $z$ multiplied by a plane
wave. That is, when
\be
m^2+1 = (1/2)^2\, , \  (3/2)^2\, , \  (5/2)^2\, ,\ldots\, 
\ee
it follows that
\be
\varphi \sim \Big(\mbox{slowly varying}\Big) \cdot\exp(i k_\mu x^\mu)\, ,
\ee
with $k_\mu \eta^{\mu\nu} k_\nu = 0$.  Hence these values of the mass parameter imply 
lightlike propagation.  Exactly this mechanism was found for all integer spin fields in $d=4$ 
cosmological backgrounds~\cite{Deser:2001pe}.

The Poincar\'e coordinate patch contains two pieces of the anti-de Sitter boundary, one at $z=0$  
and a second line at $z=\infty$ (see Figure \ref{patch}).   Demanding that solutions to the 
anti-de Sitter Klein--Gordon equation remain finite at $z=0$ requires solutions to be Bessel functions 
of the first kind, $J_\nu$, which behave as $z^\nu$ at the boundary.  At large $z$, the original 
scalar field $\varphi$ then goes as $z^{1/2}$ (up to an oscillatory factor), a potentially 
dangerous behavior at the part of the boundary with $z=\infty$.\footnote{We thank 
A.~Strominger for this important observation.}  This difficulty can be handled by studying 
wave packets with support away from $z\rightarrow\infty$.   We discuss this in more 
detail in Section~\ref{PACKETS}.

Note that to obtain the Bessel function solutions given above, we implicitly
assumed that $\omega\ne0$, that is, that $\partial_+\partial_-\varphi\ne0$.  
Additional chiral solutions, for which $\partial_+\varphi$ or $\partial_-\varphi$ vanish, 
also occur.  These pp-wave excitations will be discussed in Section~\ref{chirspec}. 
Further modified Bessel function solutions can be found by allowing $\omega$ to
become imaginary, that is, letting $k>E$~\cite{Gubser:1998bc}.  The modified 
Bessel function of the second kind decays asymptotically for large $z$, but blows up 
as $z^{-\nu}$ as $z\rightarrow 0$.  One might hope that for small masses---and
thus small values of the index $\nu$---such solutions might be admissible.  
Indeed, at the critical value $\mu=1$ these solutions behave
logarithmically, much like the finite-energy solutions found
in~\cite{Johansson}.  Unfortunately, however, in contrast to
the asymptotically anti-de Sitter behavior of those solutions,
the modified Bessel functions lead to metric components that diverge
as $z^{-2}\log z$.

Our next step is to rewrite three-dimensional topologically massive electrodynamics
and linearized gravity as scalar fields to which the above analysis applies.  More precisely, 
these ``scalars'' are indexless (nonlocal) gauge invariant components of the corresponding 
spin-1 and -2 fields.

\section{Topologically Massive AdS$_3$ Photons}

\label{gamma}

We now turn to topologically massive electrodynamics in an anti-de Sitter background,
with action
\be
I=
-\frac14\int d^3x \Big\{ \sqrt{-g}\, F_{\mu\nu} g^{\mu\rho}g^{\nu\sigma}F_{\rho\sigma}
+\mu\ \varepsilon^{\mu\nu\rho} F_{\mu\nu}A_\rho\Big\}\, .
\ee
The parameter $\mu$ has dimensions of mass.
We write $A_\mu=({A_+,A_-,A})$, where $A\equiv A_z$, and employ the light-front 
coordinates and conventions of the preceding section. Since we are dealing with local bulk 
degrees of freedom, we may assume the operator $\partial_-$ to be invertible. 
In that case, the field equation for $A_+$,
\be
\partial_- A_+ = \dot A_-  - \frac1z[z \partial + \mu + 1] \varphi
\ee
with
\be
\varphi \equiv A-\frac{\partial}{\partial_-} A_-\, ,\label{varfi}
\ee
is algebraically solvable and may be substituted back into the action, which becomes
\be
I=\int d\tau \left\{z \langle \varphi , \dot \varphi \rangle - \frac{1}{2z}
   \left([z \partial + \mu +1]\varphi\right)^2dx^- dz\right\}\, .
\ee
To obtain the light-front symplectic form, we make the field redefinition
\be
\phi = \sqrt{z} \, \varphi\, ,
\ee
so that
\be
I=\int d\tau \left\{\langle \phi,\dot\phi\rangle - \left(\frac12 [\partial \phi]^2
+\frac1{2z^2}\left[(\mu+1)^2 - \frac 14\right]\phi^2\right)dx^-dz  \right\}\, .\label{goal}
\ee
We have thus achieved our goal:  this is again the action (\ref{scalaraction}) for a scalar field, 
with mass squared
\be
\label{vecmass}
m^2 = (\mu+1)^2 -1\, .
\ee
Observe that the Bessel index for solutions to the wave equation is
\be
\nu^2 = (\mu+1)^2\, ,
\ee
so half-integer values of $\mu$ lead to lightlike propagation. The Breiten\-lohner--Freedman 
bound is saturated at $\mu=-1$.

One observation will simplify our graviton analysis in the next section.
The combination $\varphi = A -(\partial/\partial_-) A_-$ is gauge invariant under 
$\delta A= \partial \alpha$ and $\delta A_- = \partial_-\alpha$. Therefore we could have 
expedited our analysis by choosing the ``light-front gauge'' $A_-=0$.

The mass formula~\eqn{vecmass} is apparently asymmetric under 
$\mu\mapsto -\mu$. The explanation for this is an interesting interplay between the chiral 
nature of  cosmological topologically massive electrodynamics and the anti-de Sitter background. 
First, we identify 
\be
\partial_- \varphi = \partial_- A - \partial A_- = F_{-z} \equiv F_-
\ee
with the left component of the electromagnetic field strength. In flat space $F_+$, $F_-$
and $F_{+-}$ are all propagating modes with equal masses.  But our computations
show that a negative cosmological constant splits the masses of these modes:
\be
\begin{array}{c|c}
\mbox{Field} & m^2\\ \hline\hline \\[-3mm]
F_+& (\mu-1)^2-1\\[2mm]
F_{+-} & \mu^2 -1 \\[2mm]
F_{-} & (\mu+1)^2 -1
\end{array}
\label{Fasym}
\ee
Note all values of $\mu$ respect the Breiten\-lohner--Freedman
bound $m^2 \geq -1$.  These results are symmetric under a simultaneous flip of the sign of $\mu$ 
and chirality.

As a check of our computations, and to exhibit that our conclusions in no way rely on the 
invertibility of  $\partial_-$ or any other light-front peculiarity, we rederive these results
directly from the cosmological topologically massive electrodynamics equations of motion for 
the  field strengths.  Writing out the Bianchi identity
\be
\partial_{[\mu}F_{\nu\rho]}=0
\ee
in the Poincar\'e coordinates~\eqn{Poincare} yields
\be
\partial_+ F_- - \partial_- F_+ + \partial F_{+-} = 0\, ,\qquad D^\nu J_\nu = 0 \label{Bianchi}
\ee
while the equations of motion
\be
J_\nu = D^\mu F_{\mu\nu} - \frac{\mu}{2\sqrt{-g}}
   \varepsilon_\nu{}^{\mu\rho} F_{\mu\rho}=0
\ee
give
\begin{align}
\partial_+ F_- + \partial_- F_+ &=  \frac \mu z\  F_{+-}\, ,\label{one}\\[2mm]
\partial F_+ + \partial_+ F_{+-} &=  \frac{\mu-1} z\  F_+\, ,\label{two}\\[2mm]
\partial F_- - \partial _- F_{+-} &= -\frac{\mu+1} z\  F_-\, .\label{three}
\end{align}

These equations are easily manipulated to give scalar wave equations\footnote{Of 
course, all components of $F_{\mu\nu}$ can be derived from the  gauge invariant quantity 
$\varphi$ in~\eqn{varfi} (see also the Appendix).} for
any of $F_\pm$, $F_{+-}$:  for example, taking $\partial_-$ of the sum of ~\eqn{Bianchi} 
and~\eqn{one} and using~\eqn{three} to eliminate $\partial_- F_{+-}$ yields
\be
\Big[2 \partial_- \partial_+ + \partial^2 
  - \frac{(\mu+1)^2-1/4}{z^2}\Big] (\sqrt{z}F_{-})=0\, .\label{emwave}
\ee 
This is precisely the scalar wave equation of motion that follows from the scalar 
action~\eqn{goal} obtained through a light-front analysis. The same results hold for 
$F_+$ and $F_{+-}$, but with masses as quoted in~\eqn{Fasym}.

To conclude this section, we note that the energy in cosmological topologically massive 
electrodynamics is manifestly non-negative.   Since the topological Chern-Simons term 
is metric-independent, the stress-energy tensor is simply
\be
T_{\mu\nu}=-F_{\mu\rho}F_\nu{}^\rho 
    + \frac 14 g_{\mu\nu} F_{\rho\sigma} F^{\rho\sigma}\, ,
\qquad D^\mu T_{\mu\nu} = 0 \, ,
\ee
and its components are easily computed in the Poincar\'e coordinates. In particular, 
since $\partial/\partial t$ is a timelike Killing vector, the energy density becomes
$
\sqrt{-g}\, T_{0}{}^0 = \frac{z}{2} (F_+^2 + F_-^2 + F^2_{+-}) \geq 0\, .
$

\section{Topologically Massive AdS$_3$ Gravitons}

\label{gravitons}

We now come to topologically massive gravity with a cosmological 
constant. Whereas in the preceding cases the cosmological 
term provided the desired AdS$_3$ background, here it also contributes to the 
dynamics through the quadratic expansion of $\sqrt{- \det \gamma}$ 
about the $g_{\mu\nu}$ vacuum. The full action for topologically massive gravity is 
\be
I=\!\int\! d^3 x \Big\{
\!-\!\sqrt{-\gamma}\,  (R - \, 2 \Lambda)
+
\frac{1}{2\mu}\,\varepsilon^{\mu\nu\rho} \Big(\Gamma_{\mu\beta}^{\alpha}
\partial_\nu\Gamma_{\rho\alpha}^{\beta}
+\frac{2}{3}\, 
\Gamma_{\mu\gamma}^{\alpha}
\Gamma_{\nu\beta}^{\gamma}
\Gamma_{\rho\alpha}^{\beta}\Big)\Big\} .\label{TMG}
\ee
Note that we have chosen the correct, ``wrong'' sign for the $d=3$ Einstein--Hilbert 
part; as previously explained, this was necessary in the $\Lambda=0$ limit for positivity 
of energy, and remains so here. 

\vspace{.1cm}
We linearize the dynamical metric about an AdS$_3$ background by writing
\be
ds^2 = \frac{2dx^+dx^- + dz^2 }{z^2}+(dx^+)^2 h_{++} 
  + 2 dx^+ dz h_+ + dz^2 h + O(h^2)\, .
\ee
We have chosen the light-front gauge for linearized diffeomorphisms to
remove any component of the metric fluctuations $h_{\mu\nu}$ with an $x^-$ index.
Had we not done so, our computations would ultimately have depended only on the 
combination  
\be\varphi\equiv h\,  -\,  2([\partial+\frac1z]/\partial_-) h_{-z} 
  + ([\partial+\frac1z][\partial+\frac2z]/\partial_-^2) h_{--}\, ,\label{invar}\ee
which is gauge invariant under linearized diffeomorphisms 
\be\delta h_{\mu\nu}=D_\mu \xi_\nu+D_\nu \xi_\mu\, .\label{lindiffeo}\ee 

A somewhat lengthy computation yields the terms in~(\ref{TMG}) quadratic in 
the fluctuations:
\begin{gather}
I=\int d\tau \left\{\!-\frac {z^4}{\mu} \langle X, \dot h \rangle 
-\frac{z}{2}\Big( [z X - h]^2 +\frac{z^3}{\mu} \left[X+\partial h
\, +\, \frac{\mu+2}{z}\,  h\right] \, Y
\Big)dx^- dz\right\} \label{XY} \nonumber\\[1ex]
\hbox{with}\quad X\equiv \partial_- h_+\, ,\qquad Y\equiv \partial_-^2 h_{++}\, .
\end{gather}
The fields $X$ and $Y$ appear only algebraically, so can be directly integrated
out. Notice in particular that their quadratic form $X^2 + zXY/\mu$ 
is degenerate only in the pure gravity limit $\mu\rightarrow\infty$, where the  
Chern--Simons term disappears.  
This algebraic integration leaves the simple ``scalar'' action\footnote{Given 
the analysis in~\cite{Strom}, one might wonder whether integrating out the physical 
field~$h$ rather than the ``ghost''  $\partial_- h_+$ could, at least for special values of 
$\mu$, lead to an interesting ``dual'' result.  This is not the case.  Integrating out 
$h$ leads to a second order action in~$h_+$; if this action is rewritten in first order form,
using a canonical momentum~$\pi_+$ for~$\partial_-h_+$, the field $h_{++}$ becomes 
a Lagrange multiplier for a constraint that eliminates the ghost $h_+$.  The rescaling 
$\pi_+=z^{5/2}\phi/\mu$ then yields exactly the scalar action~\eqn{formaggio}, 
 with the correct physical sign.}
\be
I=\int d\tau \left\{z^3 \langle h , \dot h \rangle 
   - \frac{z}{2}\left([z \partial + \mu +3]h\right)^2dx^- dz\right\}\, .
\ee
Had we not chosen light-front gauge, but simply integrated out the auxiliary fields 
$X$ and $Y$, we would have found the same result, but with $h$ replaced by the
gauge invariant combination $\varphi$ defined in~\eqn{invar}. 
The light-front gauge choice is then simply a field redefinition.

The field rescaling
\be
\phi=z^{3/2}h
\label{scale1}
\ee
now yields our standard scalar action
\be
I=\int d\tau \left\{\langle \phi,\dot\phi\rangle - \left(\frac12 [\partial \phi]^2
+\frac1{2z^2}\left[(\mu+2)^2 - \frac 14\right]\phi^2\right)dx^-dz  \right\}\, ,
\label{formaggio}
\ee
with the mass $m$ given by
\be
\label{tensmass}
m^2 = (\mu+2)^2 -1\, .
\ee
Again, half-integer values of $\mu$ lead to lightlike propagation, with the 
Breiten\-lohner--Freed\-man bound saturated at $\mu=-2$.  Comparing  
(\ref{vecmass}) and (\ref{tensmass}), the difference between vector and tensor
masses is the simple displacement $\mu+1 \mapsto \mu+2$. 

As in electrodynamics, we find an apparently chirally  asymmetric mass formula.  
However, just as we were able to write scalar wave equations for the gauge 
invariant field strengths in  topologically massive electrodynamics, a similar 
manipulation is possible here.  We first observe that $\partial_-^2 \varphi$ 
can be written as a gauge invariant combination  
\be
\partial_-^2 \varphi = \partial_-^2h\,  -\,  2[\partial+\frac1z]\partial_- h_{-} 
  + [\partial+\frac1z][\partial+\frac2z] h_{--} = -\frac{2}{z^2}\ {\cal H}_{--}\, 
\label{Hmm}
\ee
of the metric fluctuations, where ${\cal H}_{--}$ is the chirality $-2$ component of the 
linearized cosmological Einstein tensor
\be
{\cal H}_{\rho\sigma} 
  = \left[G_{\rho\sigma}-g_{\rho\sigma}\right]_{\scriptscriptstyle\rm LINEAR}\, .
\label{linH}
\ee

It is crucial here that ${\cal H}_{\rho\sigma}$ is gauge invariant with respect to 
the linearized diffeomorphisms~\eqn{lindiffeo}:   at this order, a transformation of 
the form \eqn{lindiffeo} acting on ${\cal H}_{\rho\sigma}$ yields  the Lie derivative 
of the \emph{background} cosmological Einstein tensor 
$G_{\rho\sigma}-g_{\rho\sigma}$. This transformation therefore vanishes for an 
anti-de Sitter background.  Hence, just as in cosmological topologically massive electrodynamics, 
we describe the physical excitations in terms of these gauge invariant variables.  

In this language, the field equations are
\be
{\cal H}_{\rho\sigma}-\frac{1}{\mu}\frac{1}{\sqrt{-g}}\, 
   \varepsilon_{(\rho}{}^{\alpha\beta}D_{|\alpha|}
{\cal H}_{\sigma)\beta}=0\, ,\label{eom}
\ee
along with the Bianchi identity
\be
D^\rho {\cal H}_{\rho\sigma}=0\, .\label{bia}
\ee
{}From these equations it follows that
\be
(\Delta -\mu^2 + 3) {\cal H}_{\rho\sigma} = 0\, , \qquad  {\cal H}^\rho_\rho =0\, .
\label{wav}
\ee
Writing out equations~(\ref{eom}--\ref{wav}) in the Poincar\'e 
frame~\eqn{Poincare} yields scalar wave equations for {\it each} gauge 
invariant mode ${\cal H}_{\rho\sigma}$:
\be
\Big[2 \partial_- \partial_+ + \partial^2  - \frac{(\mu-{\rm sign}_{\rho\sigma})^2-1/4}{z^2}
  \Big] (\frac{1}{\sqrt{z}}\, {\cal H}_{\rho\sigma}) =0\, .\label{scalarwave}
\ee 
Here sign$_{\rho\sigma}$ denotes the sum of the indices $\rho + \sigma$ in light-front
coordinates (counting $z$ as $0$), and keeps track of the chirality shifts of the mass 
parameter~$\mu$.  The resulting masses for the gauge invariant modes are
\be
\begin{array}{c|c}
\mbox{Field} & m^2\\ \hline\hline \\[-3mm]
{\cal H}_{++}& (\mu-2)^2-1\\[3mm]
{\cal H}_{+z}& (\mu-1)^2-1\\[3mm]
{\cal H}_{zz}\, , \,{\cal H}_{+-} & \mu^2 -1 \\[3mm]
{\cal H}_{-z} & (\mu+1)^2 -1\\[2mm]
{\cal H}_{--} & (\mu+2)^2 -1
\end{array}
\ee
Note the (predicted)  invariance under $\mu\mapsto -\mu$ and
a chirality flip.

At this point it is clear that linearized  cosmological topologically massive gravity possesses
local ``scalar'' field theoretic degrees of freedom for any value of $\mu$, and that these
cannot be gauged away, since ${\cal H}_{\mu\nu}$ is gauge invariant. Since the mass respects 
the Breitenlohner--Freedman bound, these excitations are described by a consistent, positive 
energy field theory.   But we must still  investigate whether this theory's metric fluctuations
decay asymptotically.

\section{Asymptotics}

\label{ASYMPTOTICS}

We might next ask whether 
every solution for the curvature fluctuation~${\cal H}_{\mu\nu}$ corresponds 
to a genuine fluctuating metric.  The answer is yes, at least at this order.  In three dimensions, 
a perturbation of the Einstein tensor uniquely determines a perturbation 
of the full curvature tensor.  If ${\cal H}_{\mu\nu}$ is divergence-free with respect to the
background metric---as it is in our case---then the perturbed curvature tensor will automatically
satisfy the linearized Bianchi identities, which are the integrability conditions for the existence 
of a connection and metric.  

If we further require that these metric fluctuations produce a cosmological
Einstein tensor that vanishes asymptotically, the situation becomes more subtle.
Recall from Section \ref{HS} that the Poincar\'e patch contains two pieces of the anti-de Sitter
boundary, a surface at $z\rightarrow0$ and an additional line at $z\rightarrow\infty$.  
We start by considering $z=0$.  Bulk solutions ${\cal H}_{\rho\sigma}$ to the wave 
equation~\eqn{scalarwave} depend on $z$ through a Bessel function with index 
$\nu=\mu-{\rm sign}_{\rho\sigma}$, and are given explicitly in the Appendix.  Taking 
$\mu>0$ (say), such solutions will die off at $z=0$ for large enough $\mu$.  Indeed, when 
$0<\mu\neq1$, the asymptotics of the various components of 
the linearized cosmological Einstein tensor are  
\be
\begin{array}{c|c}
\mbox{Field} & \mbox{Asymptotics}\\ \hline\hline \\[-3mm]
{\cal H}_{++}& z^{\mu-1}\\[3mm]
{\cal H}_{+z}& z^{\mu}\\[3mm]
{\cal H}_{zz}\, , \,{\cal H}_{+-} & z^{\mu+1} \\[3mm]
{\cal H}_{-z} & z^{\mu+2}\\[2mm]
{\cal H}_{--} & z^{\mu+3}
\end{array}\label{asym}
\ee
When $\mu=1$, the Bessel function identity (valid for integer index) 
\be
J_n(x)=(-)^n J_{-n}(x)\, \label{BI}
\ee
implies  asymptotics
\be
\begin{array}{c|c}
\mbox{Field} & \mbox{Asymptotics}\\ \hline\hline \\[-3mm]
{\cal H}_{++}& z^2\\[3mm]
{\cal H}_{+z}& z\\[3mm]
{\cal H}_{zz}\, , \,{\cal H}_{+-} & z^{2} \\[3mm]
{\cal H}_{-z} & z^{3}\\[2mm]
{\cal H}_{--} & z^{4}
\end{array}\label{asym1}
\ee
where all modes decay as $z\rightarrow0$.  For negative~$\mu$, the same asymptotics 
hold but for opposite chiralities $+\leftrightarrow -$.  Only when $0<|\mu|<1$ do  
components fail to decay.

Of course, it is inconclusive to examine bare components of curvatures, since the 
unperturbed anti-de Sitter metric is singular at $z=0$ in our coordinates.  Indeed, contracting the 
curvature fluctuations in~\eqn{asym} with anti-de Sitter unit vectors gives an extra factor $z^2$, 
so no value of $\mu$ gives dangerous asymptotics for curvatures at $z=0$.  Moreover, 
invariants such as the scalar curvature and squares of the Ricci and Riemann tensors 
remain well-behaved.  For any $\mu>0$, the perturbed metric is still asymptotically 
anti-de Sitter at $z=0$, in the sense that $z^2 ds^2$ extends to the conformal boundary 
with no singularities.

In the context of the AdS/CFT correspondence, stronger conditions than the standard 
\cite{Hawking} conformal boundary conditions of general relativity are sometimes 
imposed.  It is common to express the bulk metric in Gaussian normal coordinates
\begin{align}
ds^2 
= \frac{dz^2}{z^2}  
  +\frac{1}{z^2}\Bigl[2dx^+dx^-+z^2\left({\hat h}_{++}(dx^+)^2
  +2{\hat h}_{+-}dx^+dx^-+{\hat h}_{--}(dx^-)^2\right)\Bigr]
\end{align}
and demand that the induced boundary metric satisfy the Fefferman--Graham
asymptotic conditions~\cite{Fefferman} that $({\hat h}_{++},{\hat h}_{+-},{\hat h}_{--})$ 
be finite at $z\rightarrow0$.  
It is not difficult to compute the relation between the Einstein tensor
fluctuations ${\cal H}_{\mu\nu}$ and the Gaussian normal metric ones:
\begin{align}
{\cal H}_{++}&= -\frac1{2} N(N+2){\hat h}_{++} \nonumber\\[2mm]
{\cal H}_{+-}&= \phantom{-}\frac{1}{2}N(N+2){\hat h}_{+-} \label{invert}\\[2mm]
{\cal H}_{--}&= -\frac1{2} N(N+2){\hat h}_{--} \nonumber
\end{align}
where the Euler operator \be N\equiv z\frac{\partial}{\partial z}\ee is in fact the 
unit normal vector to the anti-de Sitter boundary.  We can now easily find the asymptotic 
solutions for the Gaussian normal metric fluctuations corresponding to the curvature 
asymptotics in~\eqn{asym}:
\be
\begin{array}{c|c}
\mbox{Field} & \mbox{Asymptotics}\\ \hline\hline \\[-3mm]
{\hat h}_{++}& z^{\mu-1}\\[3mm]
{\hat h}_{+-} & z^{\mu+1} \\[3mm]
{\hat h}_{--} & z^{\mu+3}
\end{array}\label{metasym}
\ee
When $\mu>1$, the metric decays at least as fast as the Fefferman--Graham
asymptotic conditions. The value $\mu=1$ yields metric asymptotics
\be
\begin{array}{c|c}
\mbox{Field} & \mbox{Asymptotics}\\ \hline\hline \\[-3mm]
{\hat h}_{++}& z^{2}\\[3mm]
{\hat h}_{+-} & z^{2} \\[3mm]
{\hat h}_{--} & z^{4}
\end{array}\label{metasym1}
\ee
again obeying the Fefferman--Graham conditions.

The situation for large $z$, on the other hand, is more problematic.  As noted
in \cite{LSSb}, Bessel functions behave for large $z$ as
\be
J_\nu(\omega z) \sim \sqrt{\frac{2}{\pi\omega z}}
   \cos\left(\omega z - \frac{(2\nu+1)\pi}{4}\right) ,
\label{largez}
\ee  
so the curvatures $\cal H$ given in the Appendix all diverge as $z^{1/2}$, independent
of the mass parameter $\mu$.  Our linear approximation thus breaks down at large~$z$,
and in particular at the portion of the anti-de Sitter boundary contained in our Poincar\'e patch
at $z\rightarrow\infty$.

While the individual modes blow up for large $z$, however, it is important to remember
that we are working in a linearized theory, in which we can form superpositions.  We
now show that our modes can be used to construct finite norm, finite energy wave
packets with support away from large $z$.  In particular, such wave packets exist at
the critical value $\mu=1$, where they describe a propagating bulk degree of freedom
that satisfies anti-de Sitter boundary conditions.

\section{Wave Packets}

\label{PACKETS}

Superpositions of our linear solutions are most easily described in light-front gauge,
in which linearized topologically massive gravity is characterized by a single
unconstrained field $h$.  The modes we found in Section \ref{gravitons} take the form
\begin{align}
h_{\omega k}(x,z,t) &= \sqrt{\frac{\omega}{4\pi E}}\,
   \frac{1}{z}J_{\mu+2}(\omega z)e^{ikx-iEt} \\[4mm]
h^*_{\omega k}(x,z,t) &= \sqrt{\frac{\omega}{4\pi E}}\,
   \frac{1}{z}J_{\mu+2}(\omega z)e^{-ikx+iEt}\quad \ 
\hbox{with $E=\sqrt{\omega^2+k^2}$} \, .
   \nonumber
\label{hmodes}
\end{align}
The action (\ref{formaggio}) implies the existence of a conserved bilinear current
\be
{\cal J}_{\hbox{\tiny LF}}^\mu(h_1,h_2) 
  = z^4\,g^{\mu\nu}\left(h_1\partial_\nu h_2 - h_2\partial_\nu h_1\right) 
\label{sympcura}
\ee
in light-front gauge, which, as in ordinary Klein-Gordon theory, gives rise to a 
time-independent inner product
\be
(h_1,h_2) = -i\int_\Sigma dx\, dz\, z^3 h_1\overset{\leftrightarrow}{\partial}_t h^*_2
\, .
\label{proda}
\ee
Using the completeness relation for Bessel functions,
\be
\int_0^\infty dz\, z J_\nu(\omega z)J_\nu(\omega'z) 
   = \frac{1}{\omega}\delta(\omega-\omega') \, ,
\ee
it is easy to verify that 
\be
(h_{\omega k},h_{\omega' k'}) = - (h^*_{\omega k},h^*_{\omega' k'}) 
   = \delta(\omega-\omega')\delta(k-k'),  \quad (h_{\omega k},h^*_{\omega' k'}) = 0 
\, .
\ee

We can now form an arbitrary superposition
\be
h (x,z,t) = \int d\omega dk\left[ a(\omega,k)h_{\omega k}(x,z,t) 
   + a^*(\omega,k)h^*_{\omega k}(x,z,t)\right]  \, ,
\label{superpose}
\ee
which will again be a solution of the linearized field equations.  Indeed, we can 
take an \emph{arbitrary} profile $\psi(x,z)$, $\partial_t\psi(x,z)$  at $t=0$, determine 
the coefficients  $a$ and $a^*$ by
\be
a(\omega,k) = (\psi,h_{\omega k}), \qquad a^*(\omega,k) = -(\psi,h^*_{\omega k}) \, ,
\ee
and use (\ref{superpose}) to give the future evolution of the field.  In particular, 
we can choose $\psi$ to have its support away from large $z$, thus avoiding the 
asymptotic difficulties discussed at the end of Section \ref{ASYMPTOTICS}.   By
linearity of the field equations, the remaining light-front components $h_+$ and
$h_{++}$ of the metric will all have essentially the same profile, and in particular
will vanish anywhere $h$ vanishes.

The norm of $h$ is easily computed to be
\be
(h,h) = \int d\omega dk \left[ a(\omega,k)^2 - a^*(\omega,k)^2 \right] \, ,
\ee
and is preserved by time evolution.  As in ordinary Klein-Gordon theory, the norm 
is not positive definite, but, again as in ordinary Klein-Gordon theory, one can treat 
the positive- and negative-frequency modes separately to obtain a positive norm.

Although the action (\ref{formaggio}) is simply a gauge-fixed version of the full 
action of cosmological topologically massive gravity, one might still worry that 
the inner product (\ref{proda}) may be missing some of the ``gravitational'' features 
of the theory.  This is not the case.   In fact, the conserved current (\ref{sympcura}), 
and therefore the inner product, is precisely the gauge-fixed version of the full 
symplectic current of topologically massive gravity, up to a total derivative that 
gives no contribution for our wave packets.

More explicitly, any covariant Lagrangian~$L$, depending on arbitrary fields~$\varphi$, 
gives rise to a conserved symplectic current~$\cal J$ on covariant
phase space through the 
prescription \cite{Crnkovic,Lee}
\begin{align}
&\delta_1{L}[\varphi]  
  = {E}[\varphi]\delta_1\varphi 
  + \nabla_\mu{\Theta}^\mu[\varphi,\delta_1\varphi] \, , 
  \nonumber\\[4mm]
&{\cal J}^\mu[\varphi,\delta_1\varphi,\delta_2\varphi]  
     = \delta_1{\Theta}^\mu[\varphi,\delta_2\varphi] 
     -  \delta_2{\Theta}^\mu[\varphi,\delta_1\varphi]  \, ,
\label{generalsymp}
\end{align}
where the equations of motion are ${E}=0$ and $\delta_{1,2}\varphi$ satisfy 
the linearized field equations.  For cosmological topologically massive gravity, this 
current is
\begin{align}
\sqrt{-g}{\cal J}_{\hbox{\tiny TMG}}^\rho[g,\delta_1g,\delta_2g] &=
   \delta_1(\sqrt{-g}g^{\mu\nu})\delta_2\Gamma^\rho_{\mu\nu}
  - \delta_1(\sqrt{-g}g^{\rho\mu})\delta_2\Gamma^\sigma_{\mu\sigma}\\
 +&\frac{1}{2\mu}\left(
  \epsilon^{\mu\rho\sigma}\delta_1\Gamma^\alpha_{\mu\beta}
                                                      \delta_2\Gamma^\beta_{\sigma\alpha}
  - 2 \epsilon^{\rho\alpha\gamma}\delta_2g_{\mu\alpha}
                                                     {\cal H}^\mu{}_\gamma[\delta_1g]\right)
  - (1\leftrightarrow2) \, .\nonumber
\end{align}
A tedious but straightforward computation then shows that in light-front gauge,
\be 
{\cal J}_{\hbox{\tiny TMG}}^\rho[g,\delta_1g,\delta_2g]
= {\cal J}_{\hbox{\tiny LF}}^\rho[g,h_1,h_2] + \nabla_\mu{\cal F}^{[\mu\rho]} \, ,
\ee
where the superpotential ${\cal F}^{[\mu\rho]}$ makes no contribution to the 
inner product as long as the wave packets fall off fast enough at the 
boundary.\footnote{The computation is relatively easy for the $+$ and $z$ 
components of $\cal J$; one can avoid the considerably more complicated calculation 
of the $-$ component by integrating the conservation equation.}
  
We can next evaluate the energy of our wave packets.  We start again in the light-front 
formalism.  We show in the Appendix that the conserved energy takes the form
\be
H = \frac{1}{2} \int dx\,dz\, z^3\left[ (\partial_th)^2 + (\partial_xh)^2 
   + (\partial_zh)^2 + \frac{(\mu+2)^2-1}{z^2}h^2 \right] \, ,
\label{energya}
\ee
where we have used equations~(\ref{scale0}) and~(\ref{scale1}) to translate 
from $\varphi$ to $h$ and~(\ref{tensmass}) to rewrite the mass in terms of $\mu$.  
To evaluate this expression, one can use the equations of motion (\ref{scalareom}) 
and integrate by parts---again assuming that our wave packets fall off fast enough 
at the boundary---to obtain the simple expression
\be
H = \frac{1}{2} \int dx\,dz\, z^3 \left[ h\partial_t^2h - (\partial_th)^2\right] \, .
\label{energyb}
\ee
The superposition (\ref{superpose}) then gives the elegant result that
\be
H = \int d\omega dk\, \sqrt{k^2+\omega^2}\, |a(\omega,k)|^2 \, ,
\label{energyc}
\ee
which is clearly positive, and finite for suitable choices of coefficients $a(\omega,k)$.

This expression for energy was obtained from the gauge-fixed action, and one 
might again worry that it misses some ``gravitational'' features.  Again, it does not.  
For an arbitrary covariant theory, the symplectic current (\ref{generalsymp}) 
determines the symplectic structure on the covariant phase space.  Let $\Sigma$  
be any spacelike surface, with induced metric ${\tilde g}_{\mu\nu}$ and unit 
normal $n^\mu$.  Then Hamilton's equations of motion require that the 
Hamiltonian $H$ corresponding to evolution by a Killing vector $\xi^\mu$ 
(with Lie derivative ${\cal L}_\xi$) must satisfy  
\be
\delta H = \int_\Sigma d^{n-1}x\sqrt{\tilde g}\, 
    n_\mu{\cal J}^\mu[\varphi,\delta\varphi,{\cal L}_\xi\varphi] 
\label{Ham}
\ee
for any variation $\delta\varphi$ of the fields \cite{Lee}.  In particular, if 
$\delta\varphi$ is a small fluctuation around a background, the expression 
(\ref{Ham}) gives the contribution of that fluctuation to the total energy.
But we have already seen that the full symplectic current of topologically massive 
gravity is equivalent to the light-front current (\ref{sympcura}).  Substituting 
$h_2={\cal L}_th_1$ and integrating over a constant time surface, we recover 
equation (\ref{energyb}) for the energy.

We can now return to the question of asymptotics.  In the previous Section, we showed that
individual modes obeyed Fefferman--Graham asymptotics at the $z=0$ boundary, and the
preceding analysis shows that we can form wavepackets from these modes  with support only 
at the $z=0$ boundary on any time slice. However, one may worry that these wavepackets 
will propagate to the boundary at $z=\infty$ and produce non-asymptotically anti-de 
Sitter metric fluctuations there. This is not the case, as we shall now show---finite energy 
wavepackets always correspond to asymptotically anti-de Sitter metric fluctuations. 
Let us assume $\mu>0$---the
case $\mu<0$ is equivalent under a flip of chirality---and choose an initial profile 
for $h$ that has finite energy.   The energy remains constant as the wave packet 
evolves,  and since each term in the integrand of (\ref{energya}) is nonnegative, 
none can diverge.  Near the boundary at $z=0$, finiteness
requires that  $h\sim z^{-1+\varepsilon_1}$ and $\partial_\mu h\sim z^{-2+\varepsilon_2}$ with 
$\varepsilon_1,\varepsilon_2>0$.  In turn, the alternate energy formula~\eqn{energyb} and the equations of
motion for $h$
imply that $\partial_\mu\partial_\nu h \sim z^{-3+\varepsilon_3}$ ($\varepsilon_3>0$).
{}From (\ref{Hmm}) and (\ref{finalfe}),  it follows that the components 
${\cal H}_{\rho\sigma}$ of the curvature fluctuations go as $z^{-1+\delta}$
with $\delta>0$, and that the invariant components, obtained by contracting with 
anti-de Sitter unit vectors, fall off as $z^{1+\delta}$.  Moreover,  from eqn.~(\ref{invert}),
the components ${\hat h}_{\rho\sigma}$ of the metric fluctuations in Gaussian 
normal coordinates must also go as $z^{-1+\delta}$.  While this is not quite
strong enough to guarantee Fefferman-Graham asymptotics, it is sufficient to 
ensure that the spacetime remains asymptotically anti-de Sitter near $z=0$ for 
all times.

We next turn to the portion of the boundary at $z\rightarrow\infty$.  The
coordinates of our initial Poincar{\'e} patch are not well suited to describing
this region, but we can perform the inversion 
$x^\mu=-\wt x^\mu/(\wt x^\nu\eta_{\nu\rho} \wt x^\rho)$
discussed in Section~\ref{HS} to obtain new coordinates
for which this boundary occurs at ${\wt z}=0$.  This transformation takes
us out of the light front frame, and a further infinitesimal transformation is
needed to restore light front coordinates for the metric fluctuations $h$.
Once this is done, though, the energy is again of the form (\ref{energya}), with 
$x^\mu$ replaced by $\wt x^\mu$; and since $H$ is coordinate-independent,
its value is again finite (and, indeed, identical to our initial value).
By exactly the same arguments that led to asymptotically anti-de Sitter behavior at 
$z=0$, the metric must be asymptotically anti-de Sitter at $\wt z=0$.  More
generally, we can choose Poincar{\'e} coordinates near any portion of the 
boundary, and finiteness of the energy will always require asymptotically anti-de Sitter 
behavior of the excitations.

None of these considerations depend on the choice of $\mu$.  In particular, 
well-behaved, finite energy, propagating wave packets occur at the critical value 
$\mu=1$.  We stress again that these modes cannot be gauged away: they are 
expressed in terms of the gauge invariant linearized cosmological Einstein tensor.  
To be sure, these propagating configurations are not eigenfunctions of the
$\SL(2,\RR)$ generator~$L_0$ of~\cite{Strom}, so this does not \emph{directly} 
contradict their results.  But the 
suggestion of~\cite{Strom} that the theory should have no bulk modes at $\mu=1$ 
apparently misses the modes studied here. 

Although the $\mu=1$ theory has bulk modes, it does have several distinguishing features,
which we take up in the following sections.
 
\section{Chern--Simons and Chiral Gravity}

\label{CSG}

Chern--Simons formulations have proved quite useful in understanding gravity and 
supergravity in $2+1$ dimensions \cite{AT, Witten3dgrav,Hehl,Klemm,Horne,Carlipb}.   
In fact, one might adopt the motto that \emph{all} three-dimensional theories of 
gravity  are Chern--Simons.  Let us see how this is borne out for cosmological \tmg.  
Define a pair of connections
\be
{}^{\sspm}\! A_\mu{}^a{}_b 
  = \omega_\mu{}^a{}_b \pm \varepsilon^{a}{}_{bc} e_\mu{}^c\, ,
\ee
where $\omega$ and $e$ are the spin connection and dreibein.  The associated
Chern--Simons actions are
\be
I[A]= \frac{1}{2}\int d^3 x \;\varepsilon^{\mu\nu\rho} \left(A_\mu{}^a{}_b
\partial_\nu A_\rho{}^b{}_ a  + \frac{2}{3}\, 
A_\mu{}^a{}_c A_\nu{}^c{}_b A_\rho{}^b{}_a \right)\, .
\ee
If we impose the torsion-free condition as a 
constraint~\cite{DeserXiang,Carlip}  $\omega = \omega[e]$, these actions become
\begin{align}
I[{}^{\sspm}\! A[e]] 
= \int d^3 x \Biggl\{&\pm\sqrt{-\gamma}\, (R[\omega]  -2\Lambda) \\
&+ \frac12 \varepsilon^{\mu\nu\rho}
\left( \om_\mu{}^a{}_b \partial_\nu \om_\rho{}^b{}_a  + \frac{2}{3}\, 
\om_\mu{}^a{}_c \om_\nu{}^c{}_b \om_\rho{}^b{}_a \right)\Biggr\}\, .\nonumber
\end{align} 
Hence the complete cosmological \tmg\ action~(\ref{TMG}) may be written as
\be
\label{two-CS}
I_{\scriptscriptstyle\rm TMG}[e] = -\half(1-{\textstyle \frac 1\mu})I[{}^{\ssp}\! A[e]]  
  + \half(1+{\textstyle \frac 1\mu})I[{}^{\ssm}\! A[e]].
\ee
The coefficients correspond to the central charges of left and right 
components~\cite{Kraus,Solodukhin}.  
  
At the critical values of the topological 
mass parameter, 
$
\mu =  \pm 1 \, ,
$
the action reduces to
 \be
I[e]  =  - I_{\rm \scriptscriptstyle TMG}[e]\Big|_{\mu = \pm 1} 
   = \pm I[{}^{\sspm}\! A[e]]\, . \label{halfofit}
\ee 
The third order action (\ref{two-CS}) is reminiscent of the second order 
Ach\'ucarro--Town\-send--Witten~\cite{AT,Witten3dgrav} formulation of ordinary 
$d=3$ cosmological Einstein gravity as a sum of left and right $\SL(2,\RR)$ 
Chern--Simons terms,
\be
I_{\scriptscriptstyle\rm E+\Lambda}[e] = I[{}^{\ssp}\! A[e]]  - I[{}^{\ssm}\! A[e]]\, .
\ee
The model of~\eqn{halfofit} corresponds simply to discarding one of the two 
terms.\footnote{R.\ Jackiw and D.\ Grumiller (private communication) have also 
noted this.}  The theory is {\it not} topological; it still has propagating bulk modes 
that arise because of the dependence of the connection $A[e]$ on the derivatives of the
dreibein~\cite{DesJack}.  Amusingly (\ref{two-CS}) has an exact counterpart in the vector model \cite{distended}.

One can also investigate the presence of bulk modes beyond linear perturbation 
theory by examining the structure of the constraints of the full theory based on this reformulation.  
The result \cite{Carlipc,GJJ} is that the canonical analysis of~\cite{DeserXiang} for the 
case of vanishing cosmological constant generalizes in a straightforward manner, with no 
jump in the number or nature of the constraints or change in the number of degrees of freedom 
at the critical value of $\mu$.

\section{Critical Massive Gravitons as Photons}

\label{gravgamma}

We next turn to an intriguing relationship between gravity and electrodynamics, and 
show that linearized cosmological topologically massive gravity at the critical point 
$\mu=1$ reduces to topologically massive electrodynamics in the same anti-de 
Sitter background. We work in terms of the linearized cosmological Einstein
tensor ${\cal H}_{\mu\nu}$ defined in equation (\ref{linH}), and begin by observing 
that the  equations of motion~\eqn{bia} and~\eqn{wav} evaluated at a critical point 
$\mu^2=1$,
\be
(\Delta + 2) {\cal H}_{\mu\nu} = 0 = D^\mu {\cal H}_{\mu\nu} = {\cal H}^\mu_\mu\, ,
\label{harmo}
\ee
are exactly those obeyed by the metric fluctuations of pure three-dimensional
cosmological Einstein gravity in harmonic gauge.  But pure three-dimensional gravity 
has no field theoretic degrees of freedom---all its fluctuations are ``pure gauge.''  We 
can thus conclude that
\be
{\cal H}_{\mu\nu}=D_{(\mu}\wt F_{\nu)}\, ,\label{pure}
\ee
where $\wt F_\nu$ is some (suggestively labeled)
vector field.  

Our aim is therefore to reformulate linearized chiral cosmological topologically 
massive gravity in terms of the 
vector field $\wt F_\nu$.  We first write out the implications of the equation of 
motion~\eqn{eom} for $\wt F_\nu$:
\be
D_{(\mu}\Big(\wt F_{\nu)}
  -\frac1{2}\frac{1}{\sqrt{-g}}\varepsilon_{\nu)}{}^{\alpha\beta}
  D_\alpha \wt F_\beta \Big)=0\, .\label{cteom}
\ee
This equations says that the quantity 
$\wt F_{\nu}-\frac1{2}\frac{1}{\sqrt{-g}}\varepsilon_{\nu}{}^{\alpha\beta}
  D_\alpha \wt F_\beta$
vanishes modulo a Killing vector $k_\nu$, {\it i.e.},
\be
\wt F_{\nu}-\frac1{2}\frac{1}{\sqrt{-g}}\varepsilon_{\nu}{}^{\alpha\beta}
  D_\alpha \wt F_\beta=k_\nu\, ,\qquad D_{(\mu}k_{\nu)}=0\, .\label{1+curl}
\ee

At this juncture we know that any ${\cal H}_{\mu\nu}$ that obeys~\eqn{harmo}
satisfies the ansatz~\eqn{pure}, with $\wt F_\nu$ solving~\eqn{1+curl} for some
Killing vector $k_\nu$.  If we want to take $\wt F_\nu$ as our independent field, 
however, we must ask the converse question: which such $\wt F_\nu$ 
lead to linearized field strengths obeying~\eqn{harmo}?

To answer this, we start with~\eqn{1+curl}, and note that its divergence implies 
\be
D^\mu \wt F_\mu = 0.
\ee
This guarantees that ${\cal H}^\mu_\mu=0$.  

We next note that~\eqn{1+curl} can be schematically written as  
$({\mathbbm{1}} - {\rm curl})\wt F_\nu = 0$; acting with the operator 
${\mathbbm{1}}+{\rm curl}$ yields
\be
-\frac14\, (\Delta - 2) \wt F_\nu = k_\nu + \frac12\frac{1}{\sqrt{-g}} 
  \varepsilon_\nu{}^{\alpha\beta}D_\alpha k_\beta\, .
\label{killKilling}
\ee
The divergence-free condition $D^\mu {\cal H}_{\mu\nu}=0$, on the other hand,
requires that $(\Delta - 2)\wt F_\nu=0$, which is also sufficient to enforce the final 
field equation $(\Delta + 2) {\cal H}_{\mu\nu} = 0$.
We must therefore restrict \eqn{1+curl} to Killing vectors for which the right-hand
side of \eqn{killKilling} vanishes,
\be
k_\nu + \frac12\frac{1}{\sqrt{-g}} \varepsilon_\nu{}^{\alpha\beta}D_\alpha k_\beta=0\, .
\label{Killing}
\ee
We thus see that equations~\eqn{1+curl} and \eqn{Killing} together imply the full set
of field equations of linearized cosmological topologically massive gravity.  
The Killing vector $k_\nu$ is irrelevant:  we can
use \eqn{Killing} to shift $\wt F_{\nu}$ in~\eqn{1+curl}, rewriting it in the form
\be
\wt F_{\nu}+k_\nu-\frac1{2}\frac{1}{\sqrt{-g}}\varepsilon_{\nu}{}^{\alpha\beta}
  D_\alpha (\wt F_\beta+k_\beta) = 0\, .
\ee
Since the linearized field strength in \eqn{pure} depends only on the symmetric 
derivative of $\wt F_{\nu}$, any shift by a Killing vector drops out.

Finally, calling 
\be
F_{\mu\nu} = \frac{1}{\sqrt{-g}}\varepsilon_{\mu\nu\rho} (\wt F^\rho+k^\rho)
\, ,\label{dualize}
\ee
we can write our equations in the form
\be
D^\mu F_{\mu\nu}-\frac{1}{\sqrt{-g}}\varepsilon_{\nu}{}^{\mu\rho}F_{\mu\rho} 
  = 0 = D_{[\mu}F_{\nu \rho]}\, .
\ee
These are precisely the field equations for cosmological topologically massive 
electrodynamics with a mass parameter $\mu_{\scriptscriptstyle\rm EM}=2$. 
(This value is commensurate with the maximal chirality electromagnetic field strength 
being $+1$, rather than $+2$ for gravitons.)  

Our recipe is thus to solve the field equations of  topologically massive electrodynamics 
in an anti-de Sitter background for the electromagnetic field strength $F_{\mu\nu}$ or its dual 
 $\wt F_\nu$.  Such a vector determines a gauge invariant solution to linearized
cosmological topologically massive gravity.  Conversely, up to questions of  topology
and boundary conditions, any linearized solution of the gravitational theory can be 
obtained in this manner.  In other words, we have found a simple correspondence (\ref{pure}) between the single degree 
of freedom for topologically massive spin~2 and spin~1 modes at $\mu=1$.

\section{The Chiral Spectrum}

 \label{chirspec}
 
As already noted in Section~\ref{HS}, our general bulk graviton solutions can be
supplemented with additional chiral solutions of the field equations.  To study these,  
we start with a representation theoretic approach.
The AdS$_3$ background has $\SL(2,\RR)\times\SL(2,\RR)$ isometry group with Killing 
vectors
\begin{alignat}{2}
& L_+=\partial_+ \, ,\qquad &&R_+=\partial_- \, , \nonumber\\
&L_0=-x^+\partial_+-\frac12 z\partial\, , \qquad  && R_0=-x^-\partial_-
  -\frac12 z\partial\, ,\label{ISOM}\\
&L_{-}=-x^+(x^+\partial_++z\partial)+\frac12 z^2 \partial_- \, ,\qquad
  &&R_{-}=-x^-(x^-\partial_-+z\partial)+\frac12 z^2 \partial_+ \, ,\nonumber
\end{alignat}
which obey the commutation relations
\be
\begin{array}{lll}
{}[L_0,L_\pm]=\pm L_\pm\, , && [R_0,R_\pm]=\pm R_\pm\, , \\[4mm]
{}[L_+,L_-]=2L_0\, , && [R_+,R_-]=2R_0\, .
\end{array}\label{sl2sl2}
\ee
The left and right Casimirs $\Delta_L=\{L_+,L_-\}+2L_0^2$ and
$\Delta_R=\{R_+,R_-\}+2R_0^2$ are equal, and their sum yields the invariant scalar 
Laplacian
\be
\Delta = \Delta_L+\Delta_R 
  = z^2\, \Big(2\partial_+\partial_-+z\, \partial \, \frac 1z\,  \partial\Big)\, .
\ee

In the Appendix, we show that the bulk solutions we have discussed so far are not chiral
(in the three-dimensional sense), 
and review the result  that the bulk-boundary propagator is the intertwiner between 
the reducible representation of the isometry group generated by the Killing vectors and 
irreducible quasiprimary boundary fields.  There are, however, additional chiral solutions, 
obtained by requiring that $L_+$ and $R_+$ annihilate the highest chirality field 
${\cal H}_{--}$ and that $R_{+}$ annihilate all other curvature fluctuations.  In this 
case one finds discrete series representations for one of the $\SL(2,{\mathbb R})$ 
factors and a singlet for the other, and therefore a chiral subsector of the theory.

This phenomenon holds at any value of $\mu$, but is most easily exhibited at the 
critical value  $\mu=1$, where we can use the gravity/electromagnetism duality of 
Section~\ref{gravgamma}.  It is easy to verify that the field strength
\be
F=   z^{\mu-1}f(x^+) \,  dx^+\wedge dz 
   = - d\Big(\frac {z^\mu} \mu  f(x^+) \, dx^+\Big)\, ,
\ee
obeys the topologically massive Maxwell's equations in an AdS$_3$ background.
Applying the duality relations \eqn{pure} and~\eqn{dualize} at 
$\mu_{\scriptscriptstyle\rm EM}=2$, we find the curvature fluctuations
\be
{\cal H}_{++}=z^2 f'(x^+)\, ,\qquad {\cal H}_{+z}=2z f(x^+)\, .
\ee
Then using~\eqn{invert} to compute the metric fluctuations, we obtain the metric
\be
ds^2=\frac{2dx^+(dx^--\frac18z^2 f'(x^+)\, dx^+) +dz^2}{z^2}\, .
\label{ppwave}
\ee 
This is an AdS$_3$ pp-wave, and in fact solves the cosmological topologically massive 
field equations exactly, by virtue of having a vanishing cosmological Einstein tensor. 

We can also construct similar solutions at arbitrary values of $\mu$.  To that end, 
we consider the ansatz
\be
h_{++}=z^\gamma h(x^+)\, ,
\ee
and  compute the full  equations of motion for cosmological topologically massive 
gravity.  We find only a single nonvanishing component,
\be
G_{++}-g_{++}- \frac{1}{\mu\sqrt{-g}}\varepsilon_{(+}{}^{\alpha\beta}
  D_{|\alpha|}G_{+)\beta}  
  =\frac{z^\gamma h(x^+)}{2\mu}\, \gamma(\gamma+2)(\gamma-\mu+1)
  = 0\, .
\ee
Three solutions exist: $\gamma = -2$, $0$, or $\mu-1$.  The first two are also pure 
Einstein gravity solutions; of these, only $\gamma=0$ gives well-behaved Fefferman--Graham 
asymptotics.   The third is a chiral topologically massive pp-wave, the
generalization of (\ref{ppwave}) to arbitrary $\mu$; it differs
from the bulk solutions we have discussed in the previous sections.
As noted in \cite{Johansson}, at the limit $\mu\rightarrow1$ a new
logarithmic solution appears.
We also see again that $\mu=1$ is the lowest value at which the
Fefferman--Graham asymptotics are finite at the boundary, although,
as stressed in \cite{Johansson}, the logarithmic solution is still
asymptotically anti-de Sitter.

\section{Conclusions}

\label{conc}

We have analyzed cosmological topologically massive gravity,
the combination of the Einstein and Chern--Simons actions in $d=3$
with a nonvanishing cosmological constant.  
With a ``wrong'' sign Einstein term, necessary in the $\Lambda=0$ limit, the theory 
contains a single, positive energy,  massive spin-2 excitation, which can be described 
by a ``scalar.''   Indeed, the gauge invariant Einstein tensor
components themselves obey scalar wave equations with masses
that depend on their chirality. All components are related to one another by Bianchi 
identities and the field equations, and describe a single degree of freedom.
This unexpected phenomenon is reminiscent of  the
very different, massive {\em non}-gauge invariant Pauli-Fierz field in anti-de Sitter, where 
the cosmological background breaks the degeneracy of Minkowski 
states~\cite{Deser:1983tm,Deser:2001pe}.  Cosmological 
topologically massive electrodynamics behaved analogously.

With our choice of coordinates, individual Bessel function modes diverge at large 
values of $z$.  We have shown, however, that well-behaved, finite energy wave packets
exist for all values of the couplings.  These allow us to define scattering states:
we can choose a complete set of wave packets near the anti-de Sitter boundary at an early
time, allow them to evolve, and project them against another complete set of
wave packets near the boundary at a later time, without violating the requirement
that the metric be asymptotically anti-de Sitter.

We have also examined the particular, $\mu=1$,  version of 
cosmological topologically massive gravity with nonstandard sign of $G$ suggested 
in~\cite{Strom}.  This sign choice leads to negative energy for bulk modes, which  
unfortunately persist for all values of $\mu$.  Nonetheless,
we have found interesting new results at $\mu=1$. 
In particular, there is an intriguing equivalence between linearized gravity at its 
critical coupling and topologically massive electrodynamics, and a reduction from two 
to a single term in our pure Chern--Simons formulation.

It has recently been stressed in \cite{Stromc} that the theory is chiral at $\mu=1$
in the sense that the boundary conformal field theory is chiral.  Indeed, we know
from \cite{Kraus, Solodukhin} that one of the boundary central charges vanishes
at this point, and it was independently shown in \cite{Carlipc} that the
corresponding boundary term in the generator of diffeomorphisms is also zero.
Our results do not contradict this ``chirality conjecture,'' which merely implies
that our bulk modes should have a vanishing ``charge'' $E\pm J$ at the boundary.  
In particular, the extended gauge invariance discussed in \cite{Stromc} is still 
diffeomorphism invariance, albeit with one chiral sector extended all the way out to 
the boundary.  The corresponding transformations cannot remove our bulk modes,
which have nonvanishing diffeomorphism-invariant curvature fluctuations and
nontrivial curvature invariants in the bulk.

It would be interesting to investigate the boundary charges of our bulk modes further, 
but the problem is a subtle one.  The BTZ black hole exemplifies the pitfalls: the BTZ 
metric is identical for all values of $\mu$, and the shift in the boundary Virasoro 
charges cannot be seen locally, but only by considering the global diffeomorphism 
generators.  It would also be interesting to understand how our bulk modes relate
to the modes of \cite{GKP} in global coordinates, and in particular whether they 
are equivalent.

Our results are a further illustration 
of the emergent rule: adding a cosmological constant to otherwise standard actions 
gives rise to effects unsuspected from  the Minkowski perspective.

\section*{Acknowledgments}
We thank D.\ Grumiller, R.\ Jackiw, N.\ Johansson, W.\ Li, D.\ Marolf, W.\ Song, and 
A.\ Strominger for discussions. This work was supported by the National Science 
Foundation under grants 
PHY07-57190 and DMS-0636297 and by the Department of Energy under grants 
DE-FG02-91ER40674 and DE-FG02-92-ER40701. 

\medskip \medskip \medskip

\noindent {\it Note added in proof.}  Since this paper was prepared, the literature on this subject has grown considerably; see, e.g. \cite{additional-CTMG-refs}, where further references may also be found.

\newpage

\appendix

\section{Bulk Solutions and Stability}
\addtocontents{toc}{
\contentsline {section}{\\ 
Appendix: Bulk Solutions and Stability}{\thepage}}

\label{Bulk}

In this appendix we compute explicit bulk solutions for the topologically massive 
theories and examine their energies and the question of stability.  We first
return to the massive scalar wave equation,
\be
\Big[2\partial_- \partial_+ +\frac{\partial^2}{\partial z^2}
  +\frac1 z \frac{\partial}{\partial z} 
  -\frac{\nu^2}{z^2}\Big]\Big(\frac{\varphi}{z}\Big)=0\, ,\label{wave}
\ee
where the parameter $\nu$ is related to the standard scalar mass $m$ by
$$
\nu^2=m^2-1\, .
$$
As was shown in Sections~\ref{gamma} and \ref{gravitons}, the actions
for topologically massive electrodynamics and linearized topologically massive
gravity can be reduced, through nonlocal field redefinitions, to the scalar 
action, so solutions and perturbative stability results for (\ref{wave}) carry
over to our main settings.

The coordinates
\be
t=\frac{x^+-x^-}{\sqrt{2}} \quad  \mbox{and}\quad x=\frac{x^++x^-}{\sqrt2} 
\ee
determine timelike and spacelike background Killing vectors $\partial/\partial t$ and 
$\partial/\partial x$.  We diagonalize them both by expanding in Fourier modes:
\be
\varphi = \int d\omega dE\  f_{\omega}(z)a^\dagger(k,E)\exp{ (i[k x - E t])} 
  + {\rm h.c.}\, , \qquad \omega\equiv \sqrt{E^2 - k^2}\, .\label{soln}
\ee
The wave equation~\eqn{wave} implies that 
\be
f_{\omega}(z) = \frac{zJ_\nu(\omega z)}{\omega^\nu}\, .
\label{Fourier}
\ee
We assume $\nu>0$, which forces us to choose the Bessel function of the first kind 
that vanishes in the anti-de Sitter boundary, $z\rightarrow 0$. 

To compute the energy of this solution, we consider the covariantly conserved 
stress tensor
\be
T_{\mu\nu}=-\partial_\mu \varphi \partial_\nu\varphi
  +\frac12 g_{\mu\nu} \Big[\partial_\rho \varphi g^{\rho \sigma}
   \partial_\sigma \varphi + m^2 \varphi^2\Big]\, ,
\ee
omitting improvement terms that vanish on the above 
configurations.  Let us denote the timelike Killing vector by
\be
\frac{\partial}{\partial t} = \xi^\mu \partial_\mu\, ,
\ee
and recall that the vector density $\sqrt{-g} \, \xi^\mu T_\mu{}^\nu$ obeys the 
conservation equation
\be
\partial_\nu\Big(\sqrt{-g} \, \xi^\mu T_\mu{}^\nu\Big) = 0\, .
\ee
Hence the energy
\be
H = \int dx dz \sqrt{-g} \, \xi^\mu T_\mu{}^t\, ,
\ee
is conserved as long as the surface integral
\be
\int dx \sqrt{-g} \, \xi^\mu T_\mu{}^z {\Big|}_{z=0}
\ee
vanishes. An explicit computation yields
\be
\sqrt{-g} \,  \xi^\mu T_\mu{}^z \sim J_\nu(\omega z)\, 
  \Big(J_{\nu+1}(\omega z)-(\nu+1)J_{\nu}(\omega z)\Big)\sim z^{2\nu}
   \stackrel{z\rightarrow 0}{\longrightarrow}0\, .
\ee

We now compute the energy density for the configurations (\ref{soln}),
evaluated for simplicity at $t=0$ and in a frame $(E,k)=(\omega,0)$, and
find
\be
\sqrt{-g} \, \xi^\mu T_\mu{}^t = \frac{z}{2}\, 
  \Big[\omega \, J_{\nu+1}(\omega z) -\frac{\nu+1}{z}\, J_\nu(\omega z)\Big]^2
  +\frac{\nu^2 - 1}{2z}\,\left[ J_\nu(\omega z)\right]^2\, .
\ee
This is a sum of positive squares whenever $\nu^2-1 = m^2 >0$, which is the naive 
expectation for the positivity of the scalar energy.  As pointed out by Breitenlohner 
and Freedman~\cite{Breitenlohner:1982jf}, by playing the two squares off against 
one another, one need only impose the weaker condition $\nu^2 > 0$.  To see this
in our frame,
observe that the Bessel function identity
\be
-\omega J_{\nu+1}(\omega z) = [\partial -\frac{\nu}{z}] J_{\nu}(\omega z)
\ee
allows us to reexpress the energy density as
\be
\sqrt{-g} \, \xi^\mu T_\mu{}^t =\frac{z}{2}\, [\partial J_{\nu}(\omega z)]^2 
  +\frac{\nu^2}{2z}\, \left[J_\nu(\omega z)\right]^2 
  + \frac{1}{2}\, \partial[J_\nu(\omega z)]^2\, .
\ee
The first two terms are manifestly positive whenever the Breitenlohner--Freedman 
bound $\nu^2=m^2+1>0$ holds, while the final term is a vanishing surface contribution.
Our analysis closely tracks that of~\cite{Mezincescu:1984ev}, save for a different 
choice of coordinates.

To understand the representation theoretic content of these solutions, we consider 
the action of the isometry generators~\eqn{ISOM} on them.  Let us call 
$\omega_\pm\equiv (k\mp E)/\sqrt2$, so $-2\omega_+\omega_-= \omega^2$. 
Then clearly we have
\be
L_+ = i \omega_+ \,  \mbox{ and } \, R_+ = i \omega_-\, ,\label{fourier1}
\ee
where these operators are to be read as acting upon the Fourier coefficients
$a^\dagger(\omega_+,\omega_-)$.  It may be checked from (\ref{Fourier})
that the coefficients $f_\omega(z)$ satisfy the identities
\be
\Big(\omega_\pm \frac{\partial}{\partial\omega_\pm}-\frac{1}{2}z\partial
  +\frac{\nu+1}{2}\Big)\, f_\omega(z) = 0 
  = \Big[\frac{\partial}{\partial \omega_\pm}\Big(\omega_\pm
  \frac{\partial}{\partial \omega_\pm}-z\partial - 1\Big)
  +\frac12 z^2 \omega_\mp\Big]\, f_\omega(z)\, ,
\ee
which imply that $f_\omega(z)$ is highest weight.  Expressing $x^\pm$ as 
$i\partial/\partial \omega^\pm$ and giving due care to operator 
orderings, it is not difficult to determine the actions of the other isometries on the 
Fourier coefficients:
\begin{alignat}{2}
&L_0=\ \omega_+ \frac{\partial}{\partial\omega_+}-\frac{\nu-1}{2}\, ,
  && R_0=\ \omega_- \frac{\partial}{\partial\omega_-}-\frac{\nu-1}{2}\, ,
\nonumber\\[3mm]
&L_{-}=i\Big(\omega_+\frac{\partial}{\partial\omega_+}
  -\nu+1\Big)\frac{\partial}{\partial\omega_+}\, ,
  &\qquad &R_{-}=i\Big(\omega_-\frac{\partial}{\partial\omega_-}
  -\nu+1\Big)\frac{\partial}{\partial\omega_-}\, ,
  \label{fourier2}
  \end{alignat}

The operators in~\eqn{fourier1} and~\eqn{fourier2} obey the 
$\mathfrak{sl}(2,{\mathbb R})\oplus \mathfrak{sl}(2,{\mathbb R})$
Lie algebra~\eqn{sl2sl2} and are in fact closely related to the conformal 
group acting on $\mathbb R^{1,1}$.  Indeed, if we make an inverse Fourier 
transformation 
\be
a^\dagger(\omega_+,\omega_-)=\int \frac{d^2y}{(2\pi)^2} \, 
  \exp(-i [\omega_+ y^+ +\omega_- y^-])\, \chi(y^+,y^-)\label{inverse}
\, ,\ee
we find the action
\begin{alignat}{2}
& L_+=\partial_+ \, ,\qquad &&R_+=\partial_- \, , \nonumber\\[2mm]
&L_0=-y^+\partial_+ -\frac{\nu+1}2\, ,\qquad  && R_0=-y^-\partial_-
  -\frac{\nu+1}{2}\, ,\label{onshell}\\[2mm]
&L_{-}=-y^+(y^+\partial_++\nu+1)\, , \qquad
  &&R_{-}=-x^-(x^-\partial_-+\nu+1)\, ,\nonumber
\end{alignat}
on the ``boundary field'' $\chi(y^+,y^-)$. Here the 
$\SL(2,{\mathbb R})\times \SL(2,{\mathbb R})$ isometry
group acts as the two-dimensional boundary conformal group. The boundary 
field $\chi$ transforms as a weight $\nu+1$ quasiprimary.  It is not 
difficult to show that the relation between boundary and bulk fields  implied by the inverse 
Fourier transform~\eqn{inverse} and the solution~\eqn{soln} is exactly the 
bulk-boundary propagator. In detail: represent the Bessel function as 
$J_{\nu}(\omega z)=\frac{1}{2\pi i} (\frac{z}{2})^\nu
\int_C \frac{d\tau}{\tau^{\nu+1}}\exp(\tau-\frac14\omega^2 z^2\tau^{-1})$ 
(where $C$ is the Hankel contour).  This allows the integral over Fourier modes 
$\omega$ to be performed, so that 
$\varphi(z,x^\pm)=\int d^2 y \Delta(x^\pm\!-y^\pm,z) \chi(y^\pm)$, with 
the bulk-boundary propagator given in proper time representation by
$\Delta(x^\pm,z)\sim
   \int_C\frac{d\tau}{\tau^\nu}\exp(\frac\tau{z}[z^2+2x^+x^-])$. In mathematical terms, this propagator is an intertwiner
between the off-shell representation~\eqn{ISOM} and the irreducible on-shell 
one~\eqn{onshell}~\cite{intertwine}.

We next turn to topologically massive electromagnetism.  We first repeat the 
observation made in the main text that the field strength $F_{\mu\nu}$ solves 
the scalar wave equation componentwise, with masses subject to the 
Breitenlohner--Freedman bound.  Hence these modes carry positive energy 
exactly as in the scalar case.  Further,  their relation to boundary quasiprimary 
fields is also similar to the scalar case, although one must now include spin, as 
discussed in~\cite{intertwine}.   

We therefore start with a single bulk mode
\be
F_{-}=i\omega_-\, \exp(i[\omega_+ x^++\omega_- x^-]) 
  J_{\mu+1}(\omega z)+{\rm h.c.}\, ,\label{ONE}
\ee
which solves the electromagnetic wave equation~\eqn{emwave} that is, in turn, 
a consequence of the topologically massive Maxwell equations~(\ref{one}--\ref{three}) 
and the Bianchi identity~\eqn{Bianchi}.   Although we have formally reduced the 
electromagnetic action and field equations to those of a scalar, the two are not, of
course, \emph{physically} equivalent:  the new question we must address is the form 
of the  remaining components of the electromagnetic field strength.  These follow 
from the Bessel function identity
\be
\Big[\partial + \frac{\mu+1}{z}\Big]\, J_{\mu+1}(\omega z) 
   = \omega J_{\mu}(\omega z)\, .
\ee
Equation~\eqn{three} then says
\be
\partial_- F_{+-}=\Big[\partial + \frac{\mu+1}{z}\Big]\, F_- 
  = i\omega_- \exp(i[\omega_+ x^++\omega_- x^-]) \, J_{\mu}(\omega z)
  +{\rm h.c.}\, ,\label{TWO}
\ee
so
\be
F_{+-}=\exp(i[\omega_+ x^++\omega_- x^-]) J_{\mu}(\omega z)+{\rm h.c.}\, .
\ee
Similarly, equations~\eqn{one} and~\eqn{Bianchi} imply
\be
2\partial_- F_{+}=\Big[\partial + \frac{\mu}{z}\Big]\, F_{+-} 
  =  \omega\exp(i[\omega_+ x^++\omega_- x^-])\,  J_{\mu-1}(\omega z)
  +{\rm h.c.}\, ,
\ee
whence
\be
F_{+}=i\omega_+ \, 
  \exp(i[\omega_+ x^++\omega_- x^-])\,  J_{\mu-1}(\omega z)
  +{\rm h.c.}\, .\label{THREE}
\ee
We have assumed $\mu$ to be positive, and chosen the Bessel function of the first 
kind to ensure decaying behavior\footnote{Contracting with the anti-de Sitter unit 
vectors $\{ e^+=z\partial_+,\,  e^-=z\partial,\, N=z\partial\}$, all components 
$e^{+\mu}N^\nu F_{\mu\nu}$, $e^{+\mu}e^{-\nu} F_{\mu\nu}$
and $e^{-\mu}N^\nu F_{\mu\nu}$ of the field strength vanish at the boundary  
for any value of $\mu$.} for small $z$.

When the topological mass parameter takes the value 
$\mu_{\scriptscriptstyle\rm EM}=2$, we can 
employ our gravity/electromagnetism duality to construct graviton solutions. 
{}From equations~\eqn{pure} and~\eqn{dualize} and the Maxwell equations
(\ref{Bianchi},\ref{one}--\ref{three}), we can explicitly compute the relationship 
between the electromagnetic field strength and the linearized cosmological Einstein 
tensor:
\begin{align}
{\cal H}_{--}&=-z\, \partial_- F_{-z}\, ,\nn\\[2mm]
{\cal H}_{-z}\, &=-(z\partial+3)F_{-z}\, ,\nn\\[2mm]
{\cal H}_{zz}\ &=  -(z\partial +2)F_{+-}\,  = \, -2{\cal H}_{+-}\, ,\\[2mm]
{\cal H}_{+z}\, &= \, \ \; (z\partial+1)F_{+z}\, ,\nn\\[2mm]
{\cal H}_{++}&= \, \ \; z \partial_+ F_{+z}\, .\nn
\end{align}
{}From the electromagnetic solution~(\ref{ONE},\ref{TWO},\ref{THREE}), we find
\begin{align}
{\cal H}_{--}&=\phantom{-}\frac{\omega_-^2}{\omega} 
  \exp(i[\omega_+ x^++\omega_- x^-])  \ 
   z J_3(\omega z)   +{\rm h.c.} \, ,\nn\\[2mm]
{\cal H}_{-z}\, &=-i\omega_- \exp(i[\omega_+ x^++\omega_- x^-])\  z J_2(\omega z)     
  +{\rm h.c.}\, ,\nn\\[3mm]
{\cal H}_{zz}\ &=- \omega \exp(i[\omega_+ x^++\omega_- x^-])\ z J_1(\omega z) \,     
  +{\rm h.c.} \,  = \, -2{\cal H}_{+-}\, ,\label{mu1}\\[3mm]
{\cal H}_{+z}\, &= \ \ i\omega_+ \exp(i[\omega_+ x^++\omega_- x^-])\  z J_0(\omega z)   
  +{\rm h.c.} \, ,\nn\\[2mm]
{\cal H}_{++}&=- \frac{\omega_+^2}{\omega} \exp(i[\omega_+ x^++\omega_- x^-])\  
  z J_1(\omega z) +{\rm h.c.} \, .\nn
\end{align}
These solutions obey the asymptotics quoted in section~\ref{ASYMPTOTICS}.   It is 
also easy to verify that they obey the equations of motion~\eqn{eom} and Bianchi 
identity~\eqn{bia} for cosmological topologically massive gravity at $\mu=1$.  For 
arbitrary values of $\mu$, those equations read
\begin{align}
\partial_-{\cal H}_{-z}&=\Big[\partial + \frac{\mu+1}{z}\Big]\, {\cal H}_{--}\, ,\nn\\[2mm]
\partial_-{\cal H}_{zz}&=\Big[\partial + \frac\mu z\Big]\, {\cal H}_{-z}\, ,\nn\\[2mm]
2\partial_- {\cal H}_{+z}&=-\Big[\partial +\frac{\mu-1}{z}\Big]\ {\cal H}_{zz}\, ,\nn\\[2mm]
2\partial_-{\cal H}_{++}&=-\Big[\partial +\frac{\mu-2}{z}\Big]\ {\cal H}_{+z}\, ,\nn\\[2mm]
\partial_+{\cal H}_{+z}&=\Big[\partial - \frac{\mu-1}{z}\Big]\, {\cal H}_{++}\,  \label{finalfe}\\[2mm]  
\Big[\partial - \frac{1}{z}\Big]\, {\cal H}_{-z} &-\frac12 \partial_-{\cal H}_{zz}
  +\partial_+{\cal H}_{--} = 0\, ,\nn\\[2mm]
\Big[\partial -\frac 1z\Big]\, {\cal H}_{zz} &+\partial_- {\cal H}_+ +\partial_
  + {\cal H}_- = 0\, ,\nn\\[2mm]
\Big[\partial - \frac 1z\Big]\, {\cal H}_{+z} &-\frac12 \partial_+{\cal H}_{zz}
  +\partial_-{\cal H}_{++} = 0\, ,\nn
\end{align}
where we have eliminated ${\cal H}_{+-}$ in favor of ${\cal H}_{zz}$ by using
the on-shell trace condition~${\cal H}_\mu^\mu=0$.  The last three relations are the Bianchi 
identity, while the first five are the topologically massive equations of motion and Bianchi 
identity combined.

We can solve the above equations for bulk modes at arbitrary $\mu$ in much the same fashion
as for the electromagnetic case.  We find
\begin{align}
{\cal H}_{--}&=\phantom{-}\frac{\omega_-^2}{\omega} \exp(i[\omega_+ x^++\omega_- x^-])  \ 
  z J_{\mu+2}(\omega z)   +{\rm h.c.} \, ,\nn\\[2mm]
{\cal H}_{-z}\, &=-i\omega_- \exp(i[\omega_+ x^++\omega_- x^-])  \  z J_{\mu+1}(\omega z)     
  +{\rm h.c.}\, ,\nn\\[3mm]
{\cal H}_{zz}\ &=- \omega \exp(i[\omega_+ x^++\omega_- x^-])\ z J_{\mu}(\omega z) \,     
  +{\rm h.c.} \,  = \, -2{\cal H}_{+-}\, ,\label{finsols}\\[3mm]
{\cal H}_{+z}\, &= \ \ i\omega_+ \exp(i[\omega_+ x^++\omega_- x^-])\   z J_{\mu-1}(\omega z)   
  +{\rm h.c.} \, ,\nn\\[2mm]
{\cal H}_{++}&=\ \  \frac{\omega_+^2}{\omega} \exp(i[\omega_+ x^++\omega_- x^-])\  
  z J_{\mu-2}(\omega z)    +{\rm h.c.} \, .\nn
\end{align}
These results yield the asymptotics quoted in Section~\ref{ASYMPTOTICS}, and for
$\mu=1$, the expressions~\eqn{finsols} agree with those obtained via our electromagnetic 
correspondence.

{}From these curvature components, it is not difficult to invert the equations for the linearized 
Einstein tensor to obtain explicit metric fluctuations.  The light-front gauge is a particularly simple 
choice for this, but other gauges can also be used.  For example, in harmonic gauge
\be
D_\mu h^{\mu\nu} = 0, \qquad h^\mu{}_\mu=0,\label{dedo}
\ee
the metric fluctuations at $\mu=1$ become  
\begin{align}
&{h}_{zz} = -2{\tilde h}_{+-} = J_0(\omega z)e^{i(\omega_+x^+ + \omega_-x^-)}\\
&{h}_{z+} = \frac{i\omega_+}{\omega}J_1(\omega z)e^{i(\omega_+x^+ + \omega_-x^-)}\\
&{h}_{z-} 
  = \frac{i\omega_-}{\omega}\left(J_1(\omega z) 
 + \frac{2}{\omega z}J_0(\omega z)\right)e^{i(\omega_+x^+ + \omega_-x^-)}\\
&{h}_{++} 
  = \frac{\omega_+^2}{\omega^2}\left(J_0(\omega z)
  - \frac{4}{\omega z}J_1(\omega z)\right)e^{i(\omega_+x^+ + \omega_-x^-)}\\
&{h}_{--} 
  = \frac{\omega_-^2}{\omega^2}\left[\left(1 - \frac{8}{\omega^2z^2}\right)J_0(\omega z)
  - \frac{8}{\omega z}J_1(\omega z)\right]e^{i(\omega_+x^+ + \omega_-x^-)}
\end{align}


\begin{thebibliography}{99} 

\bibitem{letter} S.\ Carlip, S.\ Deser, A.\ Waldron, D.\ K.\ Wise, ``Topologically Massive AdS Gravity,'' Phys.\ Lett.\ B {\bf 666} 272-276 (2008), arXiv:0807.0486 [hep-th].

\bibitem{Deser:1982vy}
  S.~Deser, R.~Jackiw and S.~Templeton,
  ``Three-dimensional massive gauge theories,''
  Phys.\ Rev.\ Lett.\  {\bf 48}, 975 (1982);
  ``Topologically massive gauge theories,''
  Annals Phys.\  {\bf 140}, 372 (1982).

\bibitem{Deser:2003vh}
  S.~Deser and B.~Tekin,
  ``Energy in topologically massive gravity,''
  Class.\ Quant.\ Grav.\  {\bf 20}, L259 (2003);
  S.~Deser and B.~Tekin,
  ``Massive, topologically massive, models,''
  Class.\ Quant.\ Grav.\  {\bf 19}, L97 (2002),
  arXiv:hep-th/0203273.


\bibitem{Kraus}
P.~Kraus and F.~Larsen, ``Holographic gravitational anomalies,'' 
JHEP {\bf 0601}, 022 (2006), arXiv:hep-th/0508218.

\bibitem{Solodukhin} 
S.~N.~Solodukhin,
``Holography with gravitational Chern-Simons term,''
Phys.\ Rev.\ {\bf D74}, 024015 (2006), arXiv:hep-th/0509148.

\bibitem{Strom} 
 W.~Li, W.~Song, and A.~Strominger, ``Chiral gravity in three dimensions,'' 
JHEP {\bf 0804} 082 (2008), arXiv:0801.4566 [hep-th].

\bibitem{Deser:1984py}
  S.~Deser,
  ``Cosmological topological supergravity,''
  in {\it Quantum Theory of Gravity}, edited by S.~M.\ Christensen, Adam Hilger, London, 1984;
  ``Topologically massive gravity has positive energy'' (in preparation).


\bibitem{Abbott:1981ff}
  L.~F.~Abbott and S.~Deser,
  ``Stability Of Gravity With A Cosmological Constant,''
  Nucl.\ Phys.\  B {\bf 195}, 76 (1982).



\bibitem{3/2}
  S.~Deser,
  ``Massive spin 3/2 theories in three-dimensions,''
  Phys.\ Lett.\  B {\bf 140}, 321 (1984).

\bibitem{Deser:1983tm}
S.~Deser and R.~I.~Nepomechie,
``Anomalous propagation of gauge fields in conformally flat spaces,''
Phys.\ Lett.\ B {\bf 132}, 321 (1983);
``Gauge invariance versus masslessness in de Sitter space,''
Annals Phys.\ {\bf 154}, 396 (1984).




\bibitem{Deser:2001pe}
S.~Deser and A.~Waldron,
``Gauge invariances and phases of massive higher spins in (A)dS,''
Phys.\ Rev.\ Lett.\  {\bf 87}, 031601 (2001),
arXiv:hep-th/0102166;
``Partial masslessness of higher spins in (A)dS,''
Nucl.\ Phys.\ B {\bf 607}, 577 (2001),
arXiv:hep-th/0103198;
``Stability of massive cosmological gravitons,''
Phys.\ Lett.\ B {\bf 508}, 347 (2001).
arXiv:hep-th/0103255;
``Null propagation of partially massless higher spins in (A)dS and
cosmological constant speculations,''
Phys.\ Lett.\ B {\bf 513}, 137 (2001),
arXiv:hep-th/0105181.

\bibitem{Breitenlohner:1982jf}
  P.~Breitenlohner and D.~Z.~Freedman,
 ``Stability in gauged extended supergravity,''
  Annals Phys.\  {\bf 144}, 249 (1982);
  ``Positive energy in anti-de Sitter backgrounds and gauged extended
  supergravity,''
  Phys.\ Lett.\  B {\bf 115}, 197 (1982).

\bibitem{Mezincescu:1984ev}
  L.~Mezincescu and P.~K.~Townsend,
  ``Stability at a local maximum in higher dimensional anti-de Sitter space and
  applications to supergravity,''
  Annals Phys.\  {\bf 160}, 406 (1985).


\bibitem{Fefferman}
C.~Fefferman and C.R.~Graham, ``Conformal invariants,''
{\it \'Elie Cartan et les Math\'ematiques d'Aujourd'hui} (Ast\'erisque, 1985) 95.

\bibitem{Stromc} A.~Strominger,
``A simple proof of the chiral gravity conjecture,''
arXiv:0808.0506 [hep-th].


\bibitem{Moussa}
  K.~A.~Moussa, G.~Clement and C.~Leygnac,
  ``The black holes of topologically massive gravity,''
  Class.\ Quant.\ Grav.\  {\bf 20}, L277 (2003),
  arXiv:gr-qc/0303042.

\bibitem{AT} A.~Ach\'ucarro and P.K.~Townsend, ``A Chern--Simons action for 
three-dimensional anti-de Sitter supergravity theories,'' Phys.\ Lett.\ B {\bf 180}, 89 (1986).


\bibitem{Witten3dgrav} E.\ Witten, ``(2+1)-dimensional gravity as an exactly 
soluble system,'' {Nucl.\ Phys.~B} {\bf 311}, 46 (1988).

\bibitem{Gubser:1998bc}
  S.~S.~Gubser, I.~R.~Klebanov and A.~M.~Polyakov,
  ``Gauge theory correlators from non-critical string theory,''
  Phys.\ Lett.\  B {\bf 428}, 105 (1998),
  arXiv:hep-th/9802109.

\bibitem{Johansson} 
D.~Grumiller and N.~Johansson, ``Instability
in cosmological topologically massive gravity at the chiral point,'' 
JHEP {\bf 0807} 134 (2008), 
 arXiv:0805.2610 [hep-th].

\bibitem{Hawking}
S.~W.~Hawking and G.~F.~R.~Ellis, {\it The large scale structure of space-time}
(Cambridge University Press, Cambridge, 1973), section 6.8.


\bibitem{LSSb} W.~Li, W.~Song, and A.~Strominger, ``Comment on 'Cosmological 
Topological Massive Gravitons and Photons',''
 arXiv:0805.3101 [hep-th].

\bibitem{Crnkovic} C.~Crnkovi{\,c} and E.~Witten, ``Covariant description of
 canonical formalism in geometric theories,'' in {\it Three hundred years of gravitation},
 edited by S.~W.~Hawking and W.~Israel (Cambridge University Press, Cambridge, 1987).

\bibitem{Lee} J.~Lee and R.~M.~Wald, ``Local symmetries and constraints,'' 
  J.~Math.\ Phys.\ {\bf 31} (1990) 725; R.~M.~Wald, ``Black hole entropy is the Noether
  charge,'' Phys.\ Rev.\ {\bf D48} (1993) R3427.


\bibitem{Hehl} 
  P.~Baekler, E.~W.~Mielke and F.~W.~Hehl,
  ``Dynamical symmetries in topological 3-D gravity with torsion,''
  Nuovo Cim.\  B {\bf 107}, 91 (1992).

\bibitem{Klemm}  S.~Cacciatori, M.~Caldarelli, A.~Giacomini, D.~Klemm, and D.~S. Mansi, 
``Chern-Simons formulation of three-dimensional gravity with torsion and nonmetricity,''
J.\ Geom.\ Phys. {\bf 56}, 2523 (2006),  arXiv:hep-th/0507200.

\bibitem{Horne} J.~H.~Horne and E.~Witten,
  ``Conformal gravity in three dimensions as a gauge theory,''
  Phys.\ Rev.\ Lett.\  {\bf 62}, 501 (1989).

\bibitem{Carlipb} S.~Carlip,
  ``Quantum gravity in 2+1 dimensions: the case of a closed universe,''
  Living Rev.\ Rel.\  {\bf 8}, 1 (2005),
  arXiv:gr-qc/0409039.

\bibitem{DeserXiang}
S.~Deser and X.~Xiang, 
``Canonical formulations of full nonlinear topologically massive gravity,''
Phys.\ Lett.\ B {\bf 263}, 39 (1991).

\bibitem{Carlip}
S.~Carlip,
 ``Inducing Liouville theory from topologically massive gravity,''
Nucl.\ Phys.\ {\bf B362}, 111 (1991).

\bibitem{DesJack}
  S.~Deser and R.~Jackiw,
  ``Higher derivative Chern-Simons extensions,''
  Phys.\ Lett.\  B {\bf 451}, 73 (1999),
  arXiv:hep-th/9901125.

\bibitem{distended} 
S.~Deser, ``Distended Topologically Massive Electrodynamics,'' to appear in Wolfgang Kummer memorial volume, arXiv:0810.5384 [hep-th].

\bibitem{Carlipc}
S.~Carlip,
``The constraint algebra of topologically massive AdS gravity,'' 
JHEP {\bf 0810} 078 (2008)
arXiv:0807.4152 [hep-th].

\bibitem{GJJ} D.~Grumiller, R~Jackiw, and N.~Johansson,
``Canonical analysis of cosmological topologically massive gravity at the 
chiral point,'' arXiv:0806.4185 [hep-th].

\bibitem{GKP} G.~Giribet, M.~Kleban, and M.~Porrati,
``Topologically massive gravity at the chiral point is not unitary,''
JHEP {\bf 0810} 045 (2008),
arXiv:0807.4703v2 [hep-th].

\bibitem{intertwine}
V.~K.~Dobrev, 
``Intertwining operator realization of the AdS/CFT correspondence,''
Nucl. Phys. B {\bf 533}, 559 (1999), arXiv:hep-th/9812194;
S. Deser and A. Waldron,  
``Arbitrary spin representations in  de Sitter from
dS/CFT with applications to dS supergravity,''
Nucl. Phys. D {\bf 662}, 379 (2003), arXiv:hep-th/0301068.

\bibitem{additional-CTMG-refs}
G.~W.~Gibbons, C.~N.~Pope, and E.~Sezgin, 
``The General Supersymmetric Solution of Topologically Massive Supergravity,'' 
Class.\ Quant.\ Grav.\  {\bf 25}, 205005 (2008), 
arXiv:0807.2613 [hep-th].

A.~Garbarz, G.~Giribet, and Y. V\'asquez, 
``Asymptotically AdS$_3$ Solutions to Topologically Massive Gravity at Special Values of the Coupling Constants,'' 
arXiv:0811.4464 [hep-th].

M.~Henneaux, C.~Martinez, and R.~Troncoso, 
``Asymptotically anti-de Sitter spacetimes in topologically massive gravity,'' 
arXiv:0901.2874 [hep-th].

E.~A.~Bergshoeff, O.~Hohm, and P.~K.~Townsend, 
``Massive Gravity in Three Dimensions,'' 
arXiv:0901.1766 [hep-th].



\end{thebibliography}
\end{document}